\documentclass[11pt,a4paper]{article}
 
\usepackage{amsmath,amsthm,amssymb}
\usepackage{graphics,graphicx}
\usepackage{epsfig}
\usepackage{multicol}
\usepackage{color}
\makeatletter
\@addtoreset{equation}{section}
\makeatother

\usepackage{braket}

\begin{document}

\begin{flushright}
\end{flushright}
\begin{center}
\LARGE{\bf The Compton-Schwarzschild relations in higher dimensions}
\end{center}

\begin{center}
\large{\bf Matthew J. Lake} ${}^{a,b}$\footnote{matthewj@nu.ac.th}\large{\bf and Bernard Carr} ${}^{c,d}$\footnote{B.J.Carr@qmul.ac.uk} 
\end{center}
\begin{center}
\emph{ ${}^a$ The Institute for Fundamental Study, ``The Tah Poe Academia Institute", \\ 
Naresuan University, Phitsanulok 65000, Thailand \\}
\emph{ ${}^b$ Thailand Center of Excellence in Physics, Ministry of Education, Bangkok 10400, Thailand \\}
\emph{ ${}^c$ Astronomy Unit, Queen Mary University of London, Mile End Road, London E1 4NS, U.K. \\}
\emph{ ${}^d$Research Center for the Early Universe (RESCEU),
Graduate School of Science, \\ University of Tokyo, Tokyo 113-0033, Japan} 
\vspace{0.1cm}
\end{center}

\date{\today}


\begin{abstract}
In three spatial dimensions, the Compton wavelength $(R_C \propto M^{-1}$) and Schwarzschild radius $(R_S \propto M$) are dual under the transformation $M \rightarrow M_{P}^2/M$, where $M_{P}$ is the Planck mass. This suggests that there is a fundamental link -- termed the Black Hole Uncertainty Principle or Compton-Schwarzschild correspondence -- between elementary particles in the $M < M_{P}$ regime and black holes in the $M > M_{P}$ regime. In the presence of $n$ extra dimensions, compactified on some scale $R_E$, one expects $R_S \propto M^{1/(1+n)}$ for $R < R_E$, which breaks this duality. However, it may be restored in some circumstances because the effective Compton wavelength depends on the form of the $(3+n)$-dimensional wavefunction.  If this is spherically symmetric, then one still has $R_C \propto M^{-1}$, as in the $3$-dimensional case. The effective Planck length is then increased and the Planck mass reduced, allowing the possibility of TeV quantum gravity and black hole production at the LHC. However, if the wave function is pancaked in the extra dimensions and maximally asymmetric, then $R_C \propto M^{-1/(1+n)}$, so that the duality between $R_C$ and $R_S$ is preserved. In this case, the effective Planck length is reduced but the Planck mass is unchanged, so TeV quantum gravity is precluded and black holes cannot be generated in collider experiments. Nevertheless, the extra dimensions could still have consequences for the detectability of black hole evaporations and the enhancement of pair-production at accelerators on scales below $R_E$. Though phenomenologically general for higher-dimensional theories, our results are shown to be consistent with string theory via the minimum positional uncertainty derived from $D$-particle scattering amplitudes.
\end{abstract}

\section{Introduction} \label{Sec.1}

\begin{figure} \label{MR}
\begin{center}
\psfig{file=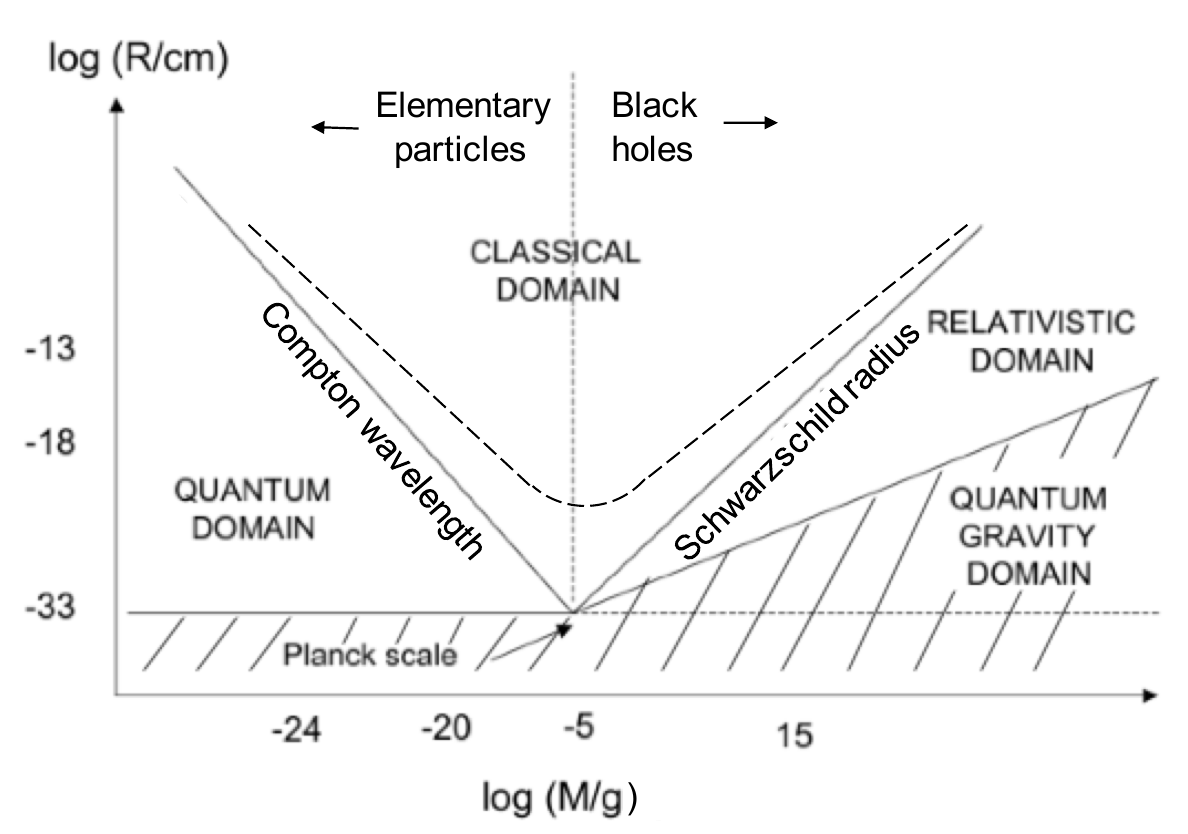,width=4.0in}
\caption{This shows the division of the ($M,R$) diagram into different physical regimes. (See Appendix A for a detailed description.) Also shown are lines corresponding to the Compton and Schwarzschild radii, and the Planck mass, length and density.}     
\end{center}
\end{figure}

A key feature of the microscopic domain is the Compton wavelength for a particle of rest mass $M$, which is $R_C = \hbar/(Mc)$. (Strictly, this is the {\it reduced} Compton wavelength, though we refer to these quanitites interchangeably throughout this paper.) In the $(M,R)$ diagram of Fig.~\ref{MR}, the region corresponding to $R<R_C$ might be regarded as the `quantum domain' in the sense that the classical description breaks down there. A key feature of the macroscopic domain is the Schwarzschild radius for a body of mass $M$, which corresponds to the size of a black hole of this mass and is $R_S = 2GM/c^2$. The region $R<R_S$ might be regarded as the `relativistic domain' in the sense that there is no stable classical configuration in this part of Fig.~\ref{MR}. 

Despite being essentially relativistic results, it is interesting that both these expressions can be derived from a semi-Newtonian treatment in which one invokes a maximum velocity $c$ but no other relativistic effects. The Compton line can be derived from the Heisenberg Uncertainty Principle (HUP), which requires that the uncertainty in the position and momentum of a particle satisfy $\Delta x \gtrsim \hbar/\Delta p$, by arguing that the momentum of a particle of mass $M$ is bounded by $Mc$. This implies that  one cannot localize it on a scale less than $\hbar/(Mc)$ and is equivalent to substituting $\Delta x \rightarrow R$ and $\Delta p \rightarrow Mc$ in the uncertainty relation. In Sec.~\ref{Sec.3}, we discuss more rigorous ways of determining the Compton scale, even in non-relativistic quantum theory, though there is always some ambiguity in the precise numerical coefficient. The expression for the Schwarzschild radius is derived rigorously from general relativity but exactly the same expression can be obtained by equating the escape velocity in Newtonian gravity to $c$. 

The Compton and Schwarzschild lines intersect at around the Planck scales,
\begin{eqnarray} \label{planck}
R_{P} = \sqrt{ \hbar G/c^3} \sim 10^{-33} \mathrm {cm} \, , \quad
M_{P} = \sqrt{ \hbar c/G} \sim 10^{-5} \mathrm g \, ,
\end{eqnarray}
and naturally divide the $(M,R)$ diagram in Fig.~\ref{MR} into three regimes, which for convenience we label quantum, relativistic and classical. (As discussed in Appendix \ref{AppendixA}, a more comprehensive discussion involves three dichotomies -- classical/quantum, non-relativistic/relativistic, weak-gravitational/strong-gravitational -- and different combinations of these then give $8$ possible regimes.) There are several other interesting lines in Fig.~\ref{MR}. The vertical line $M=M_{P}$ marks the division between elementary particles ($M <M_{P}$) and black holes ($M > M_{P}$), since the event horizon of a black hole is usually required to be larger than the Compton wavelength associated with its mass. The horizontal line $R=R_{P}$ is significant because quantum fluctuations in the metric should become important below this \cite{wheeler}.
Quantum gravity effects should also be important whenever the density exceeds the Planck value, $\rho_{P} = c^5/(G^2  \hbar) \sim 10^{94} \mathrm {g \, cm^{-3}}$, corresponding to the sorts of curvature singularities associated with the big bang or the centres of black holes \cite{CaMoPr:2011}. This implies $R < R_{P}(M/M_{P})^{1/3}$, which is well above the $R = R_{P}$ line in Fig.~\ref{MR} for $M \gg M_P$, so one might regard the shaded region as specifying the `quantum gravity' domain. This point has recently been invoked to support the notion of Planck stars \cite{rovelli} and could have important implications for the detection of evaporating black holes \cite{barrau}.   

Although the Compton and Schwarzschild boundaries correspond to straight lines in the logarithmic plot of Fig.~\ref{MR}, this form presumably breaks down near the Planck point due to quantum gravity effects. One might envisage two possibilities: either there is a smooth minimum, so the the Compton and Schwarzschild lines in some sense merge, or there is some form of phase transition  or critical point at the Planck scale, so that the separation between particles and black holes is maintained. Which alternative applies could have important implications for the relationship between elementary particles and black holes \cite{CaMuNic:2014}. 

One way of obtaining a smooth transition between the Compton and Schwarzschild lines is to invoke the Generalized Uncertainty Principle (GUP) \cite{Adler1,Adler2,Adler3}. As one approaches the Planck point from the left, this implies that the Compton wavelength is replaced by an expression of the form \cite{Ca:2014} 
\begin{equation} \label{GUP2}
R_C' =  \frac{\hbar}{Mc} + \frac{\alpha GM}{c^2} = \frac{\hbar}{Mc} \left[1  + \alpha \left( \frac{M}{M_P} \right)^2 \right]  \, ,
\end{equation}
where $\alpha$ is a dimensionless constant (normally assumed to be positive). This might  be regarded as a `generalized' Compton wavelength, the last term representing a small correction due to gravitational effects. Even if the GUP is rejected, GUP-type phenomenology can be obtained via an alternative route, by extending the de Broglie relations to super-Planckian energies \cite{Lake:2015pma,Lake:2016did}. Less attention has been paid to what happens when the black hole radius approaches the intersect point from the right. One possibility is that the Schwarzschild radius is replaced by
\begin{equation} \label{GEH1} 
R_S' = \frac{2GM}{c^2} \left[ 1 + \beta \left(\frac{M_P}{M} \right)^2 \right]
\end{equation}
for some constant $\beta$, so that the gravitational mass, which is distinct from the bare mass $M$, is given by is $M + \beta M_{P}^2/M$ \cite{CaMuNic:2014}. This is termed the Generalized Event Horizon (GEH) and represents a small perturbation of the Schwarzschild radius in the limit $M \gg M_{P}$. 

Although Eqs.~(\ref{GUP2}) and (\ref{GEH1}) apply in different regimes ($M < M_{P}$ and $M >M_{P}$, respectively), the expressions for $R_C'$ and $R_S'$ are mathematically identical if $\alpha =2$ and $\beta = 1/2$. Although  there is no reason for anticipating these values, the factor of two in the expression for the Schwarzschild radius is precise, whereas  the coefficient associated with the Compton wavelength is somewhat arbitrary, so this motivates an alternative approach in which the free constant in Eq.~(\ref{GUP2}) is associated with the first term rather than the second. It then becomes
\begin{equation} \label{GUP4A}
R_C' = \frac{\beta \hbar}{Mc} \left[1  + \frac{2}{\beta} \left(\frac{M}{M_P}\right)^2 \right]  \, ,
\end{equation}
which is mathematically identical to the expression for $R_S'$ given by Eq.~(\ref{GEH1}). What happens as one approaches the Planck point from the left and right are therefore linked.

The suggestion that there is some connection between the uncertainty principle on microscopic scales and black holes on macroscopic scales is termed the Black Hole Uncertainty Principle (BHUP) correspondence \cite{Ca:2014} and it is manifested in a unified expression for the Compton wavelength and Schwarzschild radius. We also describe this as the Compton-Schwarzschild correspondence when discussing an interpretation in terms of extended de Broglie relations \cite{Lake:2015pma,Lake:2016did}. More generally, this correspondence might allow any unified expression $R_C'(M) \equiv R_S'(M)$ which has the asymptotic behaviour $\beta \hbar/(Mc)$ for $M \ll M_{P}$ and $2GM/c^2$ for $M \gg M_{P}$. One could envisage many other unified expressions satisfying this condition but they would only be well motivated if based upon some final theory of quantum gravity. 

The BHUP correspondence suggests that  there is a different kind of positional uncertainty for an object larger than the Planck mass, related to the existence of black holes. This is not unreasonable since the Compton wavelength of a particle is smaller than the Planck length in this region,  so its physical meaning is unclear. In addition, an outside observer cannot localize an object on a scale smaller than its Schwarzschild radius. There are three important mathematical  features of the BHUP correspondence \cite{Ca:2014}. The first is the smooth transition between the Compton and Schwarzschild lines in Fig.~\ref{MR}, as indicated by the broken line. The second is the duality between the two lines under the transformation $M \rightarrow M_{P}^2/M$.  The third is the implication that there could be black holes in the sub-Planckian regime ($M <M_{P}$) with radius $\hbar/(Mc)$ for $M \ll M_{P}$. 
In this paper, we are mainly concerned with the second feature, most of our considerations being independent of the BHUP correspondence. However, our conclusions have important implications for the other two features and we discuss these elsewhere \cite{paper3}. 

The black hole boundary in Fig.~\ref{MR} assumes there are three spatial dimensions but many theories suggest that dimensionality could increase on small scales. In particular, superstring theory is consistent only in $(9+1)$ spacetime dimensions, even though our observable universe is $(3+1)$-dimensional. In current string theory models, ordinary matter is described by open strings, whose end-points are confined to a $(p+1)$-dimensional $D_p$-brane, while gravity is described by closed strings that propagate in the bulk \cite{Zwiebach:2004tj,Green:1987sp,Polchinski:1998rq}. It is therefore possible for our universe to be either a $D_3$-brane or a $D_p$-brane, which is compactified on $p-3$ extra dimensions. In either case, current experiments would be unable to directly probe the higher-dimensional nature of spacetime if the compactification scales were sufficiently small. However, there is a crucial qualitative difference between the two scenarios. In the first, matter is confined to the three visible dimensions; in the second, it can probe at least some of the extra dimensions at high energies and this more general scenario is considered in this paper. An additional motivation for this is that the $D_3$-brane of the first scenario is expected to have some thickness in quantum theory and this might be interpreted as an effective compactification scale for the confined matter.
 
This motivates us to consider the behavior of black holes and quantum mechanical particles in spacetimes with extra directions. For simplicity, we begin by assuming that all the  extra dimensions in which matter is free to propagate are compactified on a single length scale $R_E$. If there are $n$ extra dimensions, and black holes with $R_S < R_E$ are assumed to be approximately spherically symmetric with respect to the full $(3+n)$-dimensional space, then the Schwarzschild radius is given by \cite{kanti2004}
\begin{equation}
R_S =  R_E\left(\frac{M}{M_E'}\right)^{1/(1+n)} 
\label{higherBH}
\end{equation}
for $M < M_E' \equiv c^2R_E/G$, so the slope of the black hole boundary in Fig.~\ref{MR} becomes shallower. The question now arises of whether the $M$ dependence of $R_C$ is also affected by the extra dimensions. The usual assumption is that it is not, so that one still has $R_C \propto M^{-1}$. In this case, the intersect of the Schwarzschild and Compton lines is changed, so that the effective higher-dimensional Planck mass decreases (allowing the possibility of TeV quantum gravity) and the effective higher-dimensional Planck length increases  \cite{arkani}. However, in this paper we will argue that in some circumstances one expects
\begin{equation}
R_C =  R_E\left(\frac{M}{M_E}\right)^{-1/(1+n)}  
\label{higherBH}
\end{equation}
for $M > M_E \equiv \hbar/(c R_E) = M_{P}^2/M_E'$. In this case, the effective Planck length is changed but not the Planck mass, so that there can be no TeV quantum gravity regime. On the other hand, the duality between $R_C$ and $R_S$ is preserved and we will see that this has interesting physical implications. 

The plan of the paper is as follows. Sec.~\ref{Sec.2} reviews the interpretation of the Uncertainty Principle in three dimensions ($n=0$). Sec.~\ref{Sec.3} then considers how this relates to the derivation of the standard expression for the Compton wavelength. Sec.~\ref{Sec.4} discusses the (well-known) expression for the Schwarzschild radius for a ($3+n$)-dimensional black hole. Sec.~\ref{Sec.5} then derives the equivalent result for the effective Compton wavelength, emphasizing that this depends crucially on the form assumed for the wave function in the higher-dimensional space. Sec.~\ref{Sec.6} shows how the expression for the higher-dimensional Compton wavelength can be related to a `Compton temperature', which is dual to the Hawking temperature of a black hole in $3+n$ dimensions. Sec.~\ref{Sec.7} explores the consequences of our claim for the detectability of primordial black hole evaporations and recent $D$-particle scattering results. Sec.~\ref{Sec.8} gives some general conclusions and suggestions for future work. A more complete discussion of Fig.~\ref{MR} is presented in the Appendix.

\section{Interpretations of the uncertainty principle} \label{Sec.2}

In the form originally derived by Heisenberg, the uncertainty principle states that the product of the `uncertainties' in the position and momentum of a quantum mechanical particle is of order of or greater than the reduced Planck's constant $\hbar$ \cite{He27}. More generally, the rigorous definition of the uncertainty $\Delta_{\psi}O$ for an operator $\hat{O}$ is the standard deviation for a large number $N$ of (absolutely precise) repeated measurements  of an ensemble of identically prepared systems described by the wave vector $\Ket{\psi}$: 
\begin{eqnarray} \label{sd}
\Delta_{\psi}O = \sqrt{\langle \psi|\hat{O}^2|\psi\rangle - \langle \psi|\hat{O}|\psi\rangle^2} \, .
\end{eqnarray}
Formally, this expression corresponds to the limit $N \rightarrow \infty$ and is generally $|\psi\rangle$-dependent. Thus, the uncertainty $\Delta_{\psi}O$ does not correspond to incomplete knowledge about the value of the property $O$ for the system, since $\Ket{\psi}$ need not possess a definite value of $O$. 

Consistency with the Hilbert space structure of quantum mechanics requires that the product of the uncertainties associated with arbitrary operators $\hat{O}_1$ and $\hat{O}_2$ satisfy the bound  \cite{Rae00,Ish95}
\begin{eqnarray} \label{SUP}
\Delta_{\psi}O_1 \Delta_{\psi}O_2 &\geq& \frac{1}{2}\sqrt{|\langle \psi|[\hat{O}_1,\hat{O}_1]|\psi\rangle|^2 + |\langle \psi|[\hat{A},\hat{B}]_{+}|\psi\rangle|^2}
\geq \frac{1}{2}|\langle \psi|[\hat{O}_1,\hat{O}_1]|\psi\rangle| \, ,
\end{eqnarray}
where $[\hat{O}_1,\hat{O}_2]$ is the commutator of $\hat{O}_1$ and $\hat{O}_2$ and $[\hat{A},\hat{B}]_{+}$ is the anticommutator of $\hat{A} = \hat{O}_1 - \langle \hat{O}_1\rangle_{\psi}\hat{\mathbb{I}}$ and $\hat{B} = \hat{O}_2 - \langle \hat{O}_2\rangle_{\psi}\hat{\mathbb{I}}$. This formulation, which was first presented in Refs.~\cite{Ro29,Sc30}, can also be given a measurement-independent interpretation since, from a purely mathematical perspective, $\Delta_{\psi}O_1$ and $\Delta_{\psi}O_2$ represent the `widths' of the wave function in the relevant physical space or phase space, regardless of whether a measurement is actually performed.

For the operators $\hat{x}$ and $\hat{p}$, defined by $\hat{x}|\psi\rangle=x|\psi\rangle$ and $\hat{p}_x|\psi\rangle=p_x|\psi\rangle$, the commutation relation 
$[\hat{x},\hat{p}_x] = i\hbar$ gives
\begin{eqnarray} \label{SUP_xp}
\Delta_{\psi}x\Delta_{\psi}p_x \geq \hbar/2 \, ,
\end{eqnarray}
where $\Delta_{\psi}x$ and $\Delta_{\psi}p_x$ correspond to the standard deviations of $\psi(x)$ in position space and $\psi(p_x)$ in momentum space, respectively. This formulation of the uncertainty principle for $\hat{x}$ and $\hat{p}_x$ was first given in Refs.~\cite{Ke27,We28} and, for this choice of operators, the $|\psi\rangle$-dependent terms in Eq. (\ref{SUP}) are of subleading order, in accordance with Heisenberg's original result. The underlying wave-vector in the Hilbert space of the theory is identical in either the physical or momentum space representations, which correspond to different choices for the basis vectors in the expansion of $|\psi\rangle$ \cite{Rae00,Ish95}. 

Although $\Delta_{\psi}x$ and $\Delta_{\psi}p_x$ do \emph{not} refer to any unavoidable `noise', `error' or  `disturbance' introduced into the system by the measurement process, this was how Heisenberg interpreted his original result \cite{He27}. In order to distinguish between quantities representing such noise and the standard deviation of repeated measurements which do not disturb the state $\ket{\psi}$ prior to wave function collapse, we use the notation $\Delta O$ for the former and $\Delta_{\psi} O$ for the latter. 
\footnote{Strictly speaking, any disturbance to the state of the system caused by an act of measurement may also be $|\psi\rangle$-dependent. However, we adopt 
Heisenberg's original notation, in which the state-dependent nature of the disturbance is not explicit.}
In this notation, Heisenberg's original formulation of the uncertainty principle may be written as
\begin{eqnarray} \label{HUP_xp}
\Delta x\Delta p_x \gtrsim \hbar \ ,
\end{eqnarray}
ignoring numerical factors. It is well known that one can heuristically understand this result as reflecting the momentum transferred to the particle by a probing photon. However, such a statement must be viewed as a postulate, with no rigorous foundation in the underlying mathematical structure of quantum theory. Indeed, as a postulate, it has recently been shown to be manifestly false, both theoretically \cite{Oz03A,Oz03B} and experimentally \cite{Roetal12,Eretal12,Suetal13,Baetal13}. 

Despite this, the heuristic derivation of Eq. (\ref{HUP_xp}) may be found in many older texts, alongside the more rigorous derivation of Eq. (\ref{SUP}) from basic mathematical principles (see, for example, \cite{Rae00}). Unfortunately, it is not always made clear that the quantities involved in each expression are different, as clarified by the pioneering work of Ozawa \cite{Oz03A,Oz03B}. An excellent discussion of the various possible meanings and (often confused) interpretations of symbols like `$\Delta x$' is given in \cite{Scheibe}. Throughout the rest of this paper, unless explicitly stated, we consider only uncertainties of the form $\Delta_{\psi} O$, defined in Eq.~(\ref{sd}), and uncertainty relations derived from the general formula Eq.~(\ref{SUP}). 
Unfortunately, Eq.~(\ref{SUP}) is also sometimes referred to as the Generalized Uncertainty Principle or Generalized Uncertainty Relation (see, for example, \cite{Ish95}). To avoid confusion, throughout this paper we use the term {\it General} Uncertainty Principle to refer to the most general uncertainty relation obtained from the Hilbert space structure of standard non-relativistic quantum mechanics (for arbitrary operators) and the term {\it Generalized} Uncertainty Principle to refer to the amended uncertainty relation for position and momentum in non-canonical theories. 

\section{Derivations of the Compton wavelength} \label{Sec.3}

The Compton wavelength is defined as $R_C = h /(Mc)$ and first appeared historically in the expression for the Compton cross-section in the scattering of photons off electrons. Subsequently, it has arisen in various other contexts and it is important to distinguish these
when discussing how the expression for the Compton wavelength is modified in higher-dimensional models. 
The reduced Compton wavelength $\hbar /(Mc)$ appears naturally in the Klein-Gordon and Dirac equations but the non-reduced expression is relevant in processes which involve turning photon energy ($hc/\lambda$) into mass ($mc^2$). 

Other arguments
associate the Compton wavelength with the {\it localisation} of a particle and this will be relevant when we come to discuss the modifications required with extra dimensions. 
Generally speaking, the arguments are of two types, involving either the relativistic energy-momentum relation, together with the de Broglie relations, or the uncertainty principle for position and momentum from non-relativistic quantum mechanics, together with some `relativistic' arguments of a more dubious nature. We briefly review the former argument in Sec.~\ref{Sec.3.1}, for the sake of completeness, but it is the latter which chiefly concerns us in this paper and its primary deficiencies are discussed in detail in Sec.~\ref{Sec.3.2}. In Sec.~\ref{Sec.3.3}, we present an original alternative argument for identifying the maximum possible uncertainty in the momentum $(\Delta_{\psi}p_x)_{\rm max}$ with the rest mass of the particle in order to obtain a minimum value of the position uncertainty $(\Delta_{\psi}x)_{\rm min} \sim R_C$. Sec.~\ref{Sec.3.4} gives some additional comments about the relation between the Compton and Schwarzschild lines in the standard 3-dimensional case.

\subsection{ Relativistic derivation
and pair-production} \label{Sec.3.1}

The relativistic energy-momentum relation is $E^2 = P^2c^2 + M^2c^4$, where $|\vec{p}| = P = \gamma Mv$ is the magnitude of the particle's $3$-momentum, $M$ is its rest mass and $\gamma = 1/\sqrt{1-v^2/c^2}$ is the Lorentz factor. One interpretation of the Compton wavelength formula is that when $P \gtrsim \sqrt {3} Mc \; (\Rightarrow v > \sqrt{2/3} \, c)$, the particle possesses sufficient energy to pair-produce copies of itself and its antiparticle. Combining this condition with the de Broglie relation $\vec{p} = \hbar \vec{k}$ and $|\vec{k}| = 2\pi/\lambda$ gives $\lambda \lesssim h/(Mc)$, so we require $\lambda \gtrsim h/(Mc)$ to prevent this. Thus $R_C$ acts as a fundamental barrier beyond which pair-production occurs rather than further localization of the wave packet of the original particle. While it is true that the particle has sufficient energy to produce copies of itself for $P \gtrsim \sqrt {3} Mc$, it is also necessary that both the $3$-momentum in a given inertial frame and the $4$-momentum in any inertial frame are conserved  for physically allowed transitions. Additional quantum numbers may need to be conserved, depending on the particle's properties and the precise details of the quantum field theory that describes its interactions. 

An allied argument uses the fact that, in order to determine some physical property of a quantum particle, we must probe it using another kind of quantum particle. The simplest example is probing a particle of rest mass $M$ with a photon in order to determine its position. When the photon energy is greater than twice the rest energy of the particle whose position we wish to determine, $E = hc/\lambda \gtrsim 2Mc^2$, the interaction may again result in pair-production of particles of rest mass $M$, rather than further localization of the original wave packet \cite{Soviet}. The same caveats hold as before, regardless of what type of particle we probe or probe with. Likewise, it is well known that attempts to confine a particle within a radius less than its Compton wavelength result in pair-production \cite{Greiner:1990tz}, even if the energy required is derived from a confining potential (rather than a probe particle), as required for the resolution of the Klein paradox \cite{AlvarezGaume:2012zz}. Thus $R_C$ sets the distance scale at which QFT becomes essential for understanding the behavior of a particle of a given rest mass, regardless of how we localize it \cite{Baez:1999in}. 

In summary, while the de Broglie wavelength marks the scale at which non-relativistic quantum effects become important and the classical concept of a particle gives way to the idea of a wave packet, the Compton wavelength marks the point at which relativistic quantum effects become significant and the concept of a single wave packet corresponding to a state in which the particle number remains fixed becomes invalid \cite{Davies:1979mj}. $R_C$ is an effective minimum width because, on smaller scales, the concept of a single quantum mechanical particle breaks down and we must switch to a field description in which particle creation and annihilation occur in place of further spatial localization.

\subsection{Non-relativistic derivation from the uncertainty principle} \label{Sec.3.2}

If one assumes that nothing can travel faster than the speed of light, then the momentum of a particle of rest mass $M$ is bounded by $Mc$ in Newtonian theory. The uncertainty principle then implies that the particle cannot be localised on a scale less than the Compton wavelength to within a numerical factor. More precisely, if $P \lesssim Mc$, then the relations
\begin{eqnarray} \label{ident}
(\Delta_{\psi} p_x)_{\rm max} \sim M c \, , \ \ \ (\Delta_{\psi} x)_{\rm min} \sim R_C
\end{eqnarray}
yield
\begin{eqnarray} \label{Compton}
R_C \sim \frac{\hbar}{Mc} \, .
\end{eqnarray}
Since the momentum of a particle is {\it not} bounded by $Mc$ in relativity, the numerical factor is imprecise. 
Neveretheless, there are strong theoretical reasons, stemming from detailed calculations in quantum field theory, as well as compelling observational evidence \cite{Berestetsky:1982aq}, for believing that this argument is at least qualitatively correct. 

Unfortunately, the above derivation is purely heuristic and has no rigorous mathematical foundation. One problem is that the mass $M$ of a particle is a parameter in non-relativistic quantum mechanics and \emph{not} an operator, so we cannot identify it with either the expectation value $\langle\hat{O}\rangle_{\psi} = \langle \psi|\hat{O}|\psi\rangle$ or the standard deviation $\Delta_{\psi}O = \sqrt{\langle \hat{O}^2\rangle_{\psi} - \langle \hat{O}\rangle_{\psi}^2}$ of \emph{any} operator $\hat{O}$. Even if identifying $Mc$ with the expectation value of the momentum operator $\langle \hat{p}_x \rangle_{\psi}$ for a given wave packet \emph{were} permissible, there is no guarantee that the corresponding standard deviation $\Delta_{\psi}p_x$ would be of the same order of magnitude. For example, it is possible to imagine a wave packet defined by a very narrow peak, centred at $\langle \hat{p}_x \rangle_{\psi} \sim Mc$, but with spread $\Delta_{\psi}p_x \ll Mc$. However, from the general formula for $\Delta_{\psi}O$ we see that $\Delta_{\psi}O \leq \sqrt{\langle \hat{O}^2\rangle_{\psi}}$\,, which implies $(\Delta_{\psi}p_x)_{\rm max} \sim Mc$ if $\sqrt{\langle \hat{p}_x^2\rangle_{\psi}} \sim Mc$ and $\langle \hat{p}_x\rangle_{\psi} \sim 0$. This may be expected for a wave function that is symmetric, or almost symmetric, about $x=0$. 

One would expect a similar result for (almost) spherically symmetric wave packets in any number of dimensions, with $\Delta_{\psi}p_x$ being replaced by $\Delta_{\psi}P= \Delta_{\psi}|\vec{p}|$, and this scenario is considered in detail in Sec.~\ref{Sec.3.3}. We wish to place the identifications (\ref{ident}) on a firmer theoretical footing, so that we may apply the same logic to quantum systems in a higher-dimensional space with compact extra dimensions. This will enable us to derive approximate results for such a scenario \emph{without} the need for detailed QFT calculations in higher-dimensional spacetimes with compact extra dimensions.   

\subsection{An alternative argument}  \label{Sec.3.3}

In a non-relativisitic theory, the inequality $P < Mc$ may be obtained by combining the non-relativistic expression for the $3$-momentum,
\begin{eqnarray} \label{3-mom}
\vec{p} = M\vec{v} \, ,
\end{eqnarray}
with a maximum speed $|\vec{v}| < c$. In conjunction with the de Broglie relations,
\begin{equation} \label{deBroglie2}
E = \hbar\omega,  
\quad
\vec{p} = \hbar\vec{k} \, , 
\end{equation}
Eq. (\ref{3-mom}) implies 
\begin{eqnarray} \label{k^2_bound}
k = |\vec{k}| < \frac{Mc}{h}
\end{eqnarray}
and, using $k = 2\pi/\lambda$, we again recover Eq. (\ref{Compton}). However, since the speed limit is put in by hand, \emph{without} introducing additional relativistic effects, such as Lorentz invariance, the Compton limit  (\ref{Compton}) must also be inserted by hand as a lower bound on the de Broglie wavelength of the position operator eigenfunctions. Likewise, the constraint (\ref{k^2_bound}) must be imposed as an upper bound on the wavenumber of momentum operator eigenfunctions.

Mathematically, this can be achieved by defining position and momentum operators, $\hat{\vec{r}}$ and $\hat{\vec{p}}$, and their eigenfunctions in the position space representation, in the usual way,
\begin{eqnarray} \label{r}
\hat{\vec{r}} = \vec{r}, \quad
\phi(\vec{r}',\vec{r}) = \delta(\vec{r}-\vec{r}'); \quad
\hat{\vec{p}} = -i\hbar \vec{\nabla}, \quad
\phi(\vec{k},\vec{r}) = e^{i\vec{k}.\vec{r}},
\end{eqnarray} 
and then introducing an infrared cut-off in the expansion for $\psi(\hat{\vec{r}})$ in terms of $\phi(\vec{r}',\vec{r})$ or for $\psi(\vec{k})$ in terms of $\phi(\vec{k},\vec{r})$:
\begin{eqnarray} \label{r-space_exp}
\psi(\vec{r}) = \int_{h/(Mc)}^{\infty} \psi(\vec{r}')\delta(\vec{r}-\vec{r}')d^3r', \ \ \ 
\psi(\vec{k}) = \int_{h/(Mc)}^{\infty} \psi(\vec{r}')e^{i\vec{k}.\vec{r}'}d^3r' \, , 
\end{eqnarray} 
these integrals being zero for $r < h/(Mc)$. In the momentum space representation, $\hat{\vec{r}}$ and $\hat{\vec{p}}$ and their eigenfunctions take the form 
\begin{eqnarray} \label{r*}
\hat{\vec{r}} = -i\hbar \vec{\nabla}, \quad
\phi(\vec{k},\vec{r}) = e^{i\vec{k}.\vec{r}}; \quad
\hat{\vec{p}} = \vec{p},
\quad
\phi(\vec{k}',\vec{k}) = \delta(\vec{k}-\vec{k}'),
\end{eqnarray}
and consistency requires us to introduce an ultraviolet cut-off in $k$:
\begin{eqnarray} \label{k-space_exp}
\psi(\vec{r}) =  \int_{0}^{Mc/\hbar} \psi(\vec{k}')e^{-i\vec{k'}.\vec{r}}d^3k', \ \ \ 
\psi(\vec{k}) = \int_{0}^{Mc/\hbar} \psi(\vec{k}')\delta(\vec{k}-\vec{k}')d^3k'.
\end{eqnarray} 
The ultraviolet cut-off, $k_{\rm max} = Mc/\hbar$, implies an infrared cut-off, $r_{\rm min} = h/(Mc)$, and vice-versa, so that the extension of $\psi(\vec{r})$ in position space is bounded from below by the Compton wavelength, the extension of $\psi(\vec{k})$ in $k$-space is bounded from above by the corresponding wavenumber, and the extension of $\psi(\vec{p})$ in momentum-space is bounded by $P < Mc$.

Since $\Delta_{\psi} |\vec{r}|$ and $\Delta_{\psi} |\vec{p}|$  are scalars, we may write these as $\Delta_{\psi} R$ and $\Delta_{\psi} P$, respectively, where $R = |\vec{r}|$ and $P = |\vec{p}|$. For approximately spherically symmetric wave packets, we expect 
\begin{equation} 
\langle{\hat{\vec{p}}}\rangle_{\psi} \sim 0 \, , \ \ \ \Delta_{\psi}P \sim \sqrt{\langle{\hat{\vec{p}}^2}\rangle_{\psi}} \lesssim Mc \, , \label{R^2_sd}
\end{equation}
\begin{equation} 
\langle{\hat{\vec{r}}}\rangle_{\psi} \sim 0 \, , \ \ \ \Delta_{\psi}R \sim \sqrt{\langle{\hat{\vec{r}}^2}\rangle_{\psi}} \gtrsim h/(Mc). \label{P^2_sd}
\end{equation}
The commutator of $\hat{\vec{r}}$ and $\hat{\vec{p}}$ is
\begin{eqnarray} \label{Comm_R^2P^2}
[\hat{\vec{r}},\hat{\vec{p}}] = i\hbar \, ,
\end{eqnarray} 
which implies 
\begin{eqnarray} \label{SUP_R^2P^2}
\Delta_{\psi}R \, \Delta_{\psi}P \geq \hbar/2
\end{eqnarray} 
by Eq.~(\ref{SUP}). From Eqs.~(\ref{R^2_sd})-(\ref{P^2_sd}), it is therefore reasonable to make the identifications
\begin{equation} \label{P_ident}
\left(\Delta_{\psi}P\right)_{\rm max} \sim M c \, ,  \quad
\left(\Delta_{\psi}R\right)_{\rm min} \sim R_C \, , 
\end{equation}
where we will henceforth refer to the Compton wavelength as the Compton radius and restrict consideration to quasi-spherically symmetric distributions, the precise meaning of this term being explained in Sec.~\ref{Sec.5}. Under these conditions, the uncertainty relation for position and momentum \emph{does} allow us to recover the standard expression (\ref{Compton}). 

Thus, we have demonstrated that the existence of an effective cut-off for the maximum attainable energy/momentum in non-relativistic quantum mechanics implies the existence of a minimum attainable width for (almost) spherically symmetric wave functions, and this may be identified with the Compton radius for $P \lesssim Mc$. For non-spherically symmetric systems we may still consider the (maximum) upper bound on each momentum component, $p_i = \hbar k_i < Mc$, as giving rise to a (minimum) lower bound for the spatial extent of the wave packet in $i^{th}$ spatial direction. As we shall also see in Sec. \ref{Sec.5}, this has important implications for the physics of such systems in the presence of compact extra dimensions, and holds for all particles with masses $M < M_{P}$. 

The argument given above is not the usual justification for the Compton wavelength formula in non-relativistic quantum mechanics. However, the Compton radius clearly has a rigorous theoretical foundation in QFT, as well as strong empirical support, and is best thought of in a relativistic context as marking the onset of pair-production, as discussed in Sec. \ref{Sec.3.1}. Even in a non-relativistic context, in which we do not assume the cut-offs used in Eqs. (\ref{r-space_exp}) and (\ref{k-space_exp}), $\Delta_{\psi}R \lesssim h/(Mc)$ still implies $\Delta_{\psi}P \gtrsim Mc$. That is, if the width of the wavefunction in position space is less than the Compton wavelength, its width in momentum space is larger than the rest mass multiplied by the speed of light. This implies that the width of $\psi$ in energy space is of order or greater than the rest mass-energy of the particle $Mc^2$. 

Under these circumstances, a significant fraction of the measurements of $P$ in an ensemble of identically prepared systems will yield values $P \sim \Delta_{\psi}P \gtrsim Mc$, so it makes sense, even in the non-relativistic limit, to regard $R_C \sim h/(Mc)$ as a fundamental phenomenological barrier. Localizing the wave function of the particle on length scales $\Delta_{\psi}R \lesssim R_C$ ensures that the corresponding spread in momentum space, $\Delta_{\psi}P$, is sufficient to yield a significant proportion of states with energies sufficiently high to produce pairs. This is perhaps the best way to understand why the heuristic argument for the Compton wavelength formula works, even though it represents an essentially relativistic result derived in a non-relativistic theory. The advantage of the various non-relativistic arguments given above is that they can be readily extended to the higher-dimensional case with extra compactified dimensions. The results obtained are phenomenologically robust, despite being `derived' in the approximate low-energy theory. 

\subsection{Duality of Compton and Schwarzschild lines in three dimensions} \label{Sec.3.4}

The standard ($3$-dimensional) Compton and Schwarzschild lines transform into one another under both of the substitutions $M \rightarrow M_{P}^2/M$ (interchanging sub-Planckian and super-Planckian {\it mass} scales) and $R \rightarrow R_{P}^2/R$ (interchanging sub-Planckian and super-Planckian {\it length} scales). In the log-log plot of Fig.~\ref{MR}, these correspond to  reflections in the lines $M = M_{P}$ and $R = R_{P}$, respectively. There is an ambiguity in whether one interprets the Schwarzschild line as a lower bound on the localization of mass {\it outside} a black hole \cite{Ca:2014} or an upper bound on the localization of mass {\it inside} it \cite{Lake:2015pma,Lake:2016did}. In the first case, only the $M \rightarrow M_{P}^2/M$ transformation correctly preserves the direction of the associated inequalities; in the second case, only the $R \rightarrow R_{P}^2/R$ transformation does so. Note that each line maps into itself, with the upper/lower and left/right half-planes interchanging, under the combined T-duality transformation 
\begin{eqnarray} \label{T-duality}
M \rightarrow M_{P}^2/M, \ \ \ R \rightarrow R_{P}^2/R \, .
\end{eqnarray} 
T-dualities arise naturally in string theory and are known to map momentum-carrying string states to winding states and vice-versa \cite{Zwiebach:2004tj}. In addition, since they map sub-Planckian length scales to super-Planckian ones, 
switching to the dual description for $R < R_{P}$ allows the description of physical systems in an otherwise inaccessible regime \cite{Green:1987sp,Polchinski:1998rq}. Though it is unclear what role T-duality may play, at a fundamental level, in relating point particles and black holes, these considerations reinforce the suggestion that there may be such a connection.

\section{Higher-dimensional black holes}  \label{Sec.4}

The black hole boundary in Fig.~\ref{MR} assumes there are three spatial dimensions but many theories, including string theory, suggest that dimensionality could increase on sufficiently small scales \cite{Ca:2013}. 
For simplicity, we first assume that the extra dimensions are associated with a single length scale $R_E$. If the number of extra dimensions is $n$, then, in the Newtonian approximation, the gravitational force between two masses $M_1$ and $M_2$ is
\begin{equation}
F_{\mathrm{grav}} =  \frac{G_D M_1 M_2}{R^{2+n}} \, ,
 \end{equation}
where $G_D$ is the higher-dimensional gravitational constant and $D = 3+n+1$ is the number of spacetime dimensions in the relativistic theory. This becomes
\begin{equation}
F_{\mathrm{grav}} =  \frac{G M_1 M_2}{R^2} \quad \mathrm{with}  \quad G = \left(\frac{G_D}{R_E^n}\right) 
\label{newG}
 \end{equation}
for $R \gtrsim R_E$, so one recovers the inverse-square law there. The higher-dimensional nature of the gravitational force is only manifest for $R \lesssim R_E$. This follows directly from the fact that general relativity can be extended to an arbitrary number of dimensions, so we may take the weak field limit of Einstein's field equations in $3+n+1$ dimensions to obtain the Newtonian gravitational potential generated by a mass $M$ as \cite{Maartens:2010ar,Maartens:2005ww}
\begin{equation} \label{V1}
V_{\mathrm{grav}} =  \frac{G_D M}{R^{1+n}} \, 
 \end{equation}
 for $R \lesssim R_E$. This becomes
\begin{equation} \label{V2}
V_{\mathrm{grav}} =  \frac{G M}{R}  
 \end{equation}
for $R \gtrsim R_E$. In the Newtonian limit, the effective gravitational constants at large and small scales are different because of the dilution effect of the extra dimensions. 

There are two interesting mass scales associated with the length scale $R_E$: the mass whose Compton wavelength is $R_E$, 
\begin{eqnarray}  \label{M_E} 
M_{E} \sim \frac{\hbar}{c R_E} \sim  M_{P}\frac{R_{P}}{R_E} \, ,
\label{ME}
\end{eqnarray} 
and the mass whose Schwarzschild radius is $R_E$,
\begin{eqnarray}  \label{M_C} 
M_E' \sim \frac{M_{P}^2}{M_E} \sim M_{P} \frac{R_E}{R_{P}} \,  .
\label{ME'}
\end{eqnarray}  
These mass scales are reflections of each other in the line $M=M_{P}$. An important implication of Eq.~(\ref{V1}) is that the usual expression for the Schwarzschild radius no longer applies for masses below $M_E'$. If the black hole is assumed to be (approximately) spherically symmetric in the higher-dimensional space on scales $R \ll R_E$, the expression for $R_S$ must be replaced with
\begin{equation} \label{higherBH}
R_S \sim R_E  \left(\frac{M}{M_E'} \right)^{1/(n+1)} \sim R_{*}\left(\frac{M}{M_{P}}\right)^{1/(1+n)},
 \end{equation}
where
\begin{equation} \label{R*}
R_{*} \sim (R_{P}R_E^n)^{1/(1+n)}.
\end{equation}
Therefore, the slope of the black hole boundary in Fig.~\ref{MR} becomes shallower for $M \lesssim M_E'$.

Strictly speaking, the metric associated with Eq.~(\ref{higherBH}) is only valid for infinite extra dimensions, since it assumes asymptotic flatness \cite{Hor12}. For black hole solutions  with compact extra dimensions, one must ensure periodic boundary conditions with respect to the compact space. However, Eq.~(\ref{higherBH}) should be accurate for black holes with $R_S \ll R_E$, so we adopt this for the entire range $R_{P} \lesssim R \lesssim R_E$ as a first approximation. Similar problems arise, even in the Newtonian limit, since Eq.~(\ref{V2}) is also only valid for infinite extra dimensions and does not respect the periodicity of the internal space. In practice, we expect any corrections to smooth out the line around $R_S \sim R_E$, so that the true metric yields the asymptotic forms corresponding to the Schwarzschild radius of a $(3+1)$-dimensional black hole on scales $R_S \gg R_E$ and a $(3+n+1)$-dimensional black hole on scales $R_S \ll R_E$. 

This behavior is indicated in Fig.~2(a) for various values of $n$. The intersect with the Compton boundary (assuming this is unchanged) then becomes
\begin{equation} \label{revisedplanck}
R_{P}' 
\sim (R_{P}^2R_E^n)^{1/(2+n)}, \quad   
M_{P}' 
\sim (M_{P}^2M_E^n)^{1/(2+n)} \,.
 \end{equation}
This gives $M_{P}' \sim M_{P}$ and $R_{P}' \sim R_{P}$ for $R_E \sim R_{P}$ but $M_{P}'  \ll M_{P}$ and $R_{P}'  \gg R_{P}$ for $R_E  \gg R_{P}$. In principle, $M_{P}'$  could be of order $1$~TeV, making it accessible by the Large Hadron Collider (LHC). According to standard arguments \cite{arkani}, this would allow quantum gravity effects to be detectable in accelerator experiments, providing 
\begin{equation} \label{nconstraint}
R_E  \sim 10^{(32/n) - 17} \mathrm{cm}
\sim  \left \lbrace
\begin{array}{rl}
&10^{15} \quad (n=1) \\
&10^{-1} \quad (n=2) \\
&10^{-13} \quad (n=7) \\
&10^{-17} \quad (n=\infty) \, .
\end{array}\right.
  \end{equation}
Clearly, $n=1$ is excluded on empirical grounds but $n=2$ is possible. One expects $n=7$ in M-theory \cite{Becker:2007zj}, so it is interesting that $R_E$ is of order a Fermi if all of these dimensions extend beyond the Planck scale. $R_E \rightarrow 10^{-17}$cm as $n \rightarrow \infty$ since this is the smallest scale which can be probed by the LHC. 

The above analysis assumes that all the extra dimensions have the same size. One could also consider a hierarchy of compactification scales, $R_i = \alpha_i R_{P}$ with $\alpha_1 \geq \alpha_2 \geq .... \geq \alpha_n \geq 1$, such that the dimensionality progressively increases as one goes to smaller distances \cite{Ca:2013}. In this case, the effective {\it average} length scale associated with the compact internal space is
\begin{equation} 
\langle R_E \rangle = \left( \prod_{i=1}^{n} R_i  \right)^{1/n} = R_{P} \left( \prod_{i=1}^{n} \alpha_i \right)^{1/n} \, .
\end{equation} 
and the new effective Planck scales are 
\begin{equation}
R_{P}' \sim  \left(R_{P}^2 \prod_{i=1}^{n} R_i\right)^{1/(2+n)} \sim (R_{P}^2 \langle R_E \rangle^n)^{1/(2+n)}  
\end{equation}
\begin{equation}
M_{P}' \sim \left(M_{P}^2 \prod_{i=1}^{n}M_i^n\right)^{1/(2+n)} \sim (M_{P}^2 \langle M_E \rangle ^{n})^{1/(2+n)} \, ,
\label{revisedplanck*}
\end{equation}
where $M_i \sim \hbar/(cR_i)$ and $\langle M_E \rangle \sim \hbar/(c \langle R_E \rangle )$. For $R_{k+1} \lesssim R \lesssim R_{k}$, the effective Schwarzschild radius is then given by 
\begin{equation} \label{effectiveschwarz}
R_S = R_{*(k)}\left(\frac{M}{M_{P}}\right)^{1/(1+k)}, \ \ \ R_{*(k)} = \left(R_{P} \prod_{i=1}^{k \leq n}R_{i}\right)^{1/(1+k)}.
\end{equation}
This situation is represented in Fig.~2(b). Clearly, the Planck scales are not changed as much in this case
as in the scenario for which the $n$ extra dimensions all have the same scale. 


\begin{figure}  \label{MR2}
\begin{center}
   \psfig{file=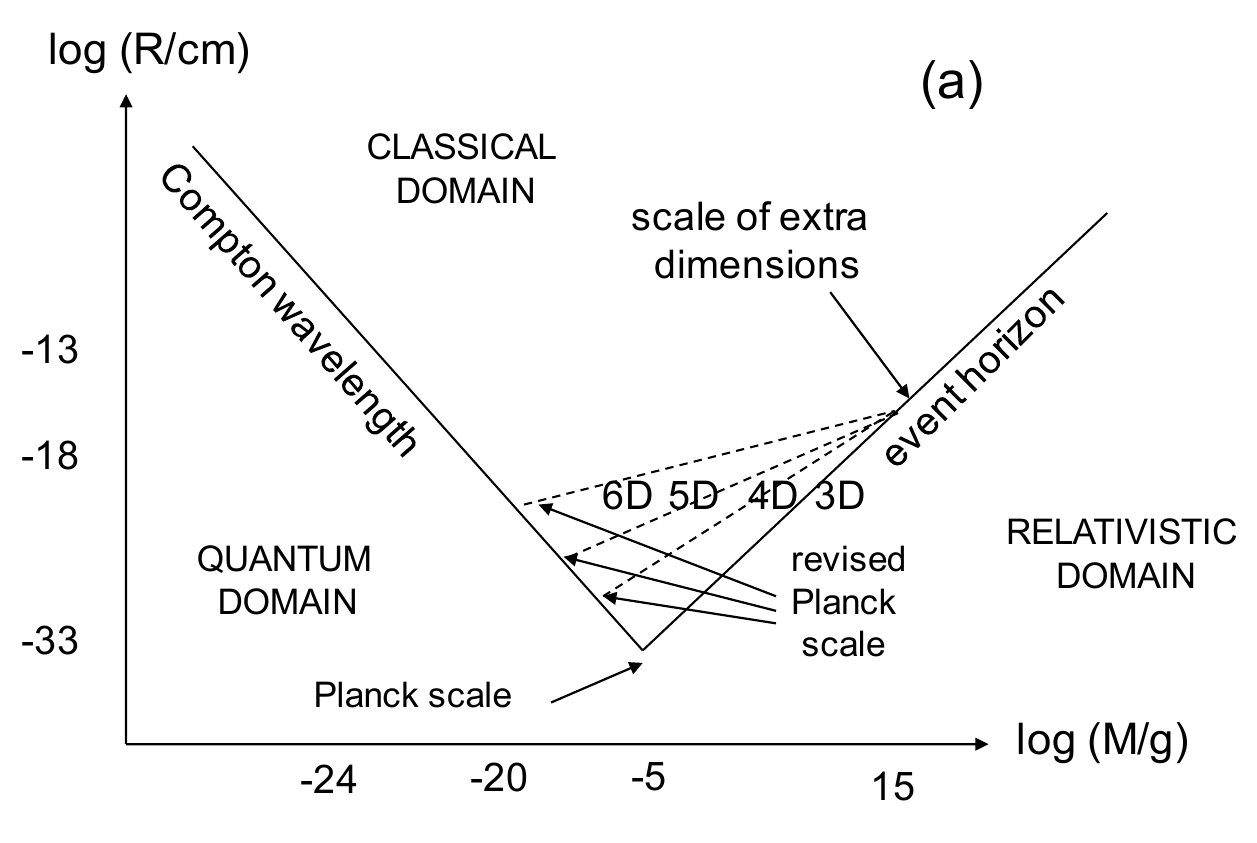,width=3.5in}
   \psfig{file=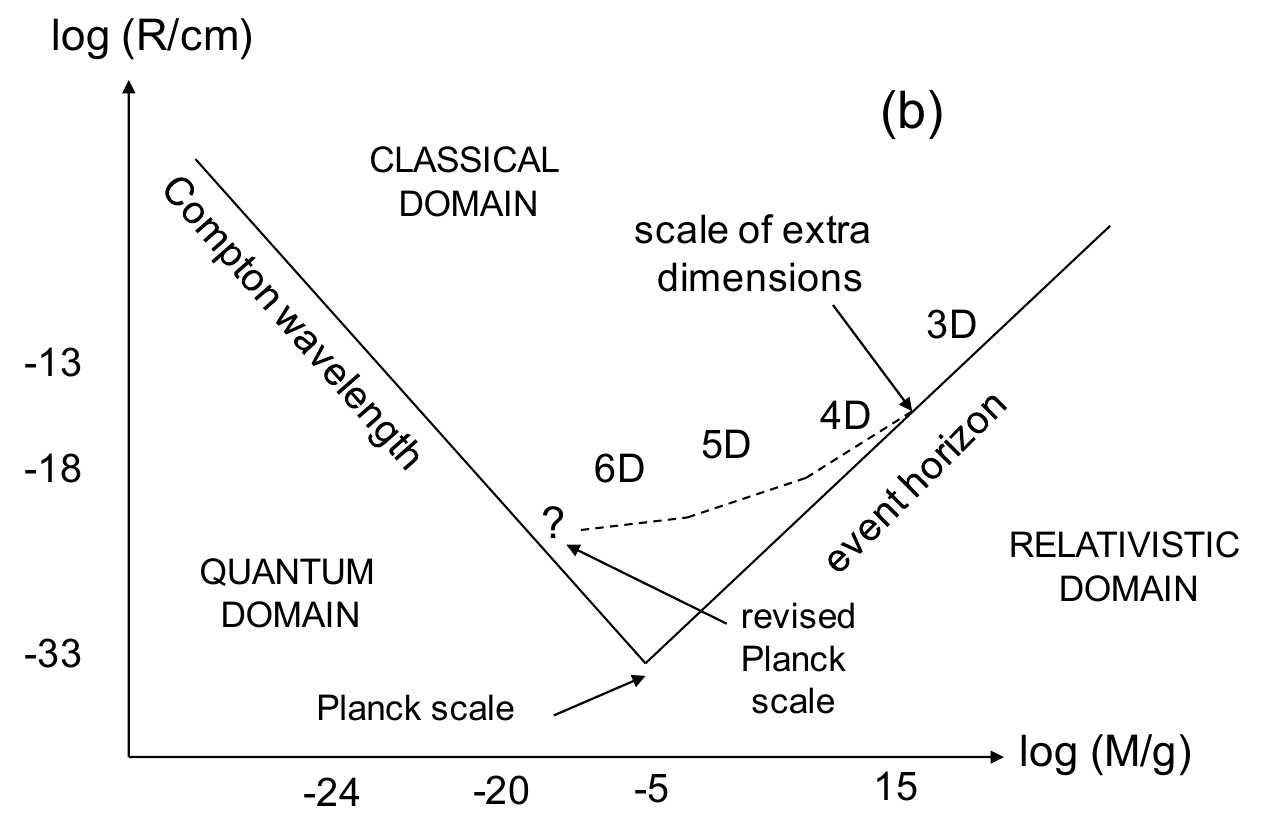,width=3.5in}
  \caption {Modification of the Schwarzschild line in the $(M,R)$ diagram in the presence of extra compact dimensions associated with a single length scale (a) or a hierarchy of length scales (b). If the Compton scale preserves its usual form, the effective Planck scales are shifted as indicated.} 
\end{center}
\end{figure} 

\begin{figure}  \label{Fig7}
\begin{center}
   \psfig{file=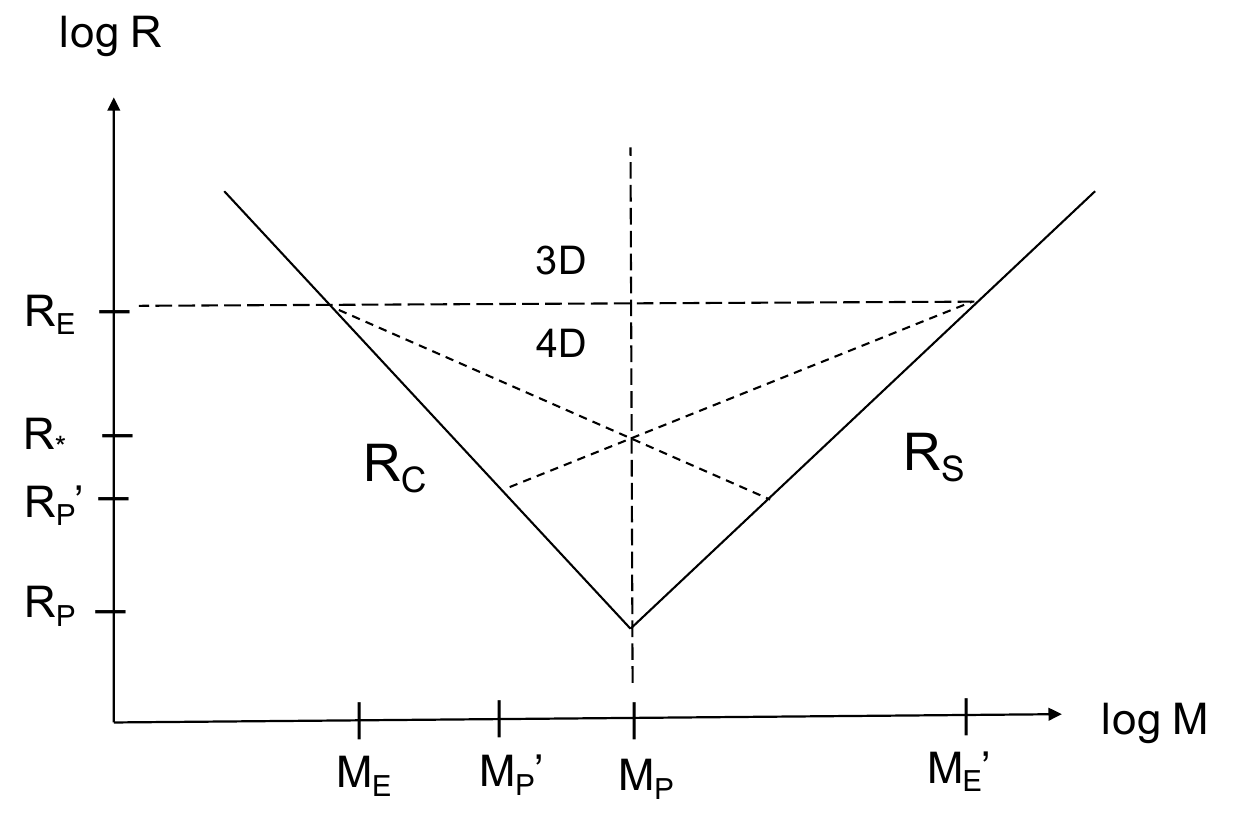,width=3.5in}
  \caption {Key mass and length scales in the 3D case (solid lines) and 4D cases (dotted lines) if the extra dimension is compactified on a scale $R_E$. The associated Compton and Schwarzschild masses are $M_E$ and $M_E'$, respectively. The revised Planck scales are $M_{P}'$ and $R_{P}'$ if duality is violated but $M_{P}$ and $R_*$ if it is preserved.}  
\end{center}
\end{figure}

The relationship between the various key scales ($R_E, R_E',R_{P},R_{P}',M_{P},M_{P}',R_*$) in the above analysis is illustrated in Fig.~\ref{Fig7} for the case of one extra spatial dimension ($n=1$). This suggests that the duality between the Compton and Schwarzschild length scales is lost  if one introduces extra spatial dimensions. However, this raises the issue of whether the expression for the standard Compton wavelength should also be modified in the higher-dimensional case and we now address this. We argue that, in this scenario, a phenomenologically important length scale is the {\it effective} Compton wavelength, which may be identified with the minimum effective width (in $3$-dimensional space) of the higher-dimensional wave packet $(\Delta_{\psi}x)_{\rm min}$.

\section{Quantum particles in higher dimensions} \label{Sec.5}

In this section, we consider whether the effective Compton wavelength in a ($3+n + 1$)-dimensional spacetime with $n$ compact dimensions scales like $M^{-1}$, as in the $3$-dimensional case, or according to a different scaling law. If we use the term in the sense discussed in Sec.~3, to describe the localisabiity of a particle, we find that this depends crucially on the degree to which the wave packet of the particle is `pancaked' in the extra dimensions, i.e. on the degree of asymmetry between its size in the infinite and compact dimensions.

\subsection{Uncertainty Principle in higher dimensions} \label{Sec.5.1}

In 3-dimensional space with Cartesian coordinates $(x,y,z)$, the uncertainty relations for position and momentum are
\begin{equation} 
\Delta_{\psi}x\Delta_{\psi}p_{x} \gtrsim \hbar \, , \quad
\Delta_{\psi}y\Delta_{\psi}p_{y} \gtrsim \hbar \, , \quad
\Delta_{\psi}z\Delta_{\psi}p_{z} \gtrsim \hbar \label{3D_URz} \, .
\end{equation}
For spherically symmetric distributions, we have
\begin{eqnarray}  \label{SpherSymm3D-1}
\Delta_{\psi}x \sim \Delta_{\psi}y \sim \Delta_{\psi}z \sim \Delta_{\psi}R \, , \quad
\Delta_{\psi}p_{x} \sim \Delta_{\psi}p_{y} \sim \Delta_{\psi}p_{z} \sim \Delta_{\psi}P \, ,
\end{eqnarray}
where the axes are arbitrarily orientated, so that the relations (\ref{3D_URz}) are each equivalent to
\begin{eqnarray}  \label{3D_UR_SpherSymm} 
\Delta_{\psi}R \, \Delta_{\psi}P \gtrsim c R_{P}M_{P} = \hbar \, .
\end{eqnarray}
In $(3+n)$ spatial dimensions, we also have
\begin{eqnarray} \label{(3+n)-D_UR}
\Delta_{\psi}x_i \, \Delta_{\psi}p_{i} \gtrsim \hbar \quad (i = 1,2, . . . n) \, ,
\end{eqnarray}
so that for distributions that are spherically symmetric with respect to the three large dimensions we obtain 
\begin{eqnarray}  \label{(3+n)-D_UR_combined} 
\Delta_{\psi}R \, \Delta_{\psi}P \left(\prod_{i=1}^{n} \Delta_{\psi} \, x_i\Delta_{\psi}p_{i}\right)  \gtrsim  \hbar^{1+n} \, .
\end{eqnarray}
The exponent on the right is $1+n$, rather than $3+n$, because there is only {\it one} independent relation associated with the large spatial dimensions due to spherical symmetry. Assuming, for simplicity, that the extra dimensions are compactified on a single length scale $R_E$, then spherically symmetric wave functions are only possible on scales $\Delta_{\psi}R < R_E$ in position space or $\Delta_{\psi}P > c M_E$ in momentum space. In this case, we may identify the standard deviations in the extra dimensions (i.e. in both position  and momentum space) with those in the infinite dimensions, 
\begin{eqnarray}  \label{SpherSymm(3+n)D-1} 
\Delta_{\psi}x_{i} \sim \Delta_{\psi}R \, , \quad
\Delta_{\psi}p_{i} \sim \Delta_{\psi}P \, , 
\end{eqnarray}
for all $i$, so that Eq.~(\ref{(3+n)-D_UR_combined}) reduces to (\ref{3D_UR_SpherSymm}). Following the usual identifications, this gives the standard expression for the Compton wavelength in a higher-dimensional context. 

However, this is not the only possibility. The condition of spherical symmetry in the three large dimensions implies that the directly observable part of $\psi$ is characterized by a single length scale, the 3-dimensional radius of the wave packet $\Delta_{\psi}\tilde{R}$. One can therefore consider states for which the  wave function occupies a ($1+n$)-dimensional volume and define a length scale $\Delta_{\psi}R$ as the \emph{average} radius across these $1+n$ dimensions: 
\begin{eqnarray} \label{Volume}
V_{(1+n)} \sim \Delta_{\psi}\tilde{R} \prod_{i=1}^{n}\Delta_{\psi}x_i \sim (\Delta_{\psi}R)^{1+n} \, ,
\end{eqnarray}
This volume is of special interest because it is the only one that can be constructed from all $1+n$ {\it independent} length scales associated with the wave packet. 
\\
\indent
In this scenario, $\Delta_{\psi}\tilde{R} \neq \Delta_{\psi}R$ and $\Delta_{\psi}x_i \neq \Delta_{\psi}R$, for at least some $i$. Such states may be considered `quasi-spherical' in the sense that they are spherically symmetric with respect to the three large dimensions but pancake-shaped (and possibly extremely irregular) from the higher-dimensional perspective. They are also degenerate with fully spherically symmetric wave packets of radius $\Delta_{\psi}R$, in that they are associated with the same length scale. In this case, we have
 \begin{eqnarray} \label{(3+n)-D_UR*1}
(\Delta_{\psi}R)^{1+n}\Delta_{\psi}P \left(\prod_{i=1}^{n} \Delta_{\psi}p_{i}\right) \gtrsim  \hbar^{1+n} \, .
\end{eqnarray}
where $\Delta_{\psi}P \equiv \Delta_{\psi}\tilde{P}$ denotes the spread of the wave function in the three infinite dimensions of momentum space. This relation comes from combining the {\it single} independent 3-dimensional uncertainty relation, associated with the large dimensions, with the $n$ independent uncertainty relations associated with the compact dimensions. With respect to the full $(3+n)$-dimensional space, the wave packet need not be spherically symmetric.

Let us now restrict ourselves to states for which
\begin{eqnarray}  \label{SpherSymm(3+n)D-2**} 
\Delta_{\psi}p_{i} \sim \kappa_{i}^{-1} cM_{P} \, , 
\end{eqnarray} 
where the $\kappa_i$ are dimensionless constants satisfying
\begin{eqnarray} \label{X}
1 \leq \kappa_i \leq \frac{R_E}{R_{P}} = \frac{M_{P}}{M_E} \, .
\end{eqnarray} 
This ensures that 
\begin{eqnarray} \label{P_range}
cM_{P} \geq \Delta_{\psi}p_i \geq cM_E
\end{eqnarray} 
and restricts us to the higher-dimensional region of the $(M,R)$ diagram. Conditions (\ref{(3+n)-D_UR}) then reduce to
\begin{eqnarray} \label{(3+n)-D_UR*}
\Delta_{\psi}x_i \gtrsim \kappa_i R_{P} \, ,
\end{eqnarray}
which together with Eq. (\ref{X}) ensures
\begin{eqnarray}
R_{P} \leq (\Delta_{\psi}x_i)_{\rm min} \leq R_E \, .
\end{eqnarray} 
Note that the $\kappa_i$ have no intrinsic relation with the constants $\alpha_i$ used to characterize the hierarchy of length scales in Sec.~\ref{Sec.4}, and we are still considering the case in which all extra dimensions are compactified on a single length scale $R_E$. They  simply paramaterize the degree to which each extra dimension is `filled' by the wave packet (e.g. if $\kappa_i = 1$, the physical spread of the wave packet in the $i^{th}$ extra dimension is $R_{P}$). Were we to consider a similar parameterization in the hierarchical case, it would follow immediately that $\kappa_i \leq \alpha_i$.

Equation (\ref{(3+n)-D_UR*1}) now becomes
\begin{eqnarray}  \label{(3+n)-D_UR_combined**} 
\Delta_{\psi}R \gtrsim R_{P}\left[\frac{cM_{P}(\prod_{i=1}^{n}\kappa_{i})}{\Delta_{\psi}P}\right]^{1/(1+n)} \, .
\end{eqnarray}
The validity of this bound is subject to the quasi-spherical symmetry condition (\ref{Volume}) but it is \emph{stronger} than the equivalent condition (\ref{3D_UR_SpherSymm}) for fully spherically symmetric states. By definition, such a wave packet is also quasi-spherically symmetric in momentum space, in the sense that it is spherically symmetric with respect to three infinite momentum dimensions, but not with respect to the full $(3+n)$-dimensional momentum space. 

For fully spherically symmetric states, we must put each $\kappa_i$ equal to a single value $\kappa$, where $\Delta_{\psi}R \sim \kappa R_{P}$ and $\Delta_{\psi}P \sim \kappa^{-1} cM_{P}$ for the 3-dimensional part of the wave function because the momentum space representation $\psi(P)$ is given by the Fourier transform of $\psi(R)$. Hence, a wave function that is totally spherically symmetric in the $3+n$ dimensions of position space will also be totally spherically symmetric in the $3+n$ dimensions of momentum space. Therefore, considering spherically symmetric wave functions with $R_C \sim (\Delta_{\psi}R)_{\rm min}$ and $(\Delta_{\psi}P)_{\rm max} \sim cM$ (corresponding to $\kappa = M_{P}/M$) just restores the standard Compton formula in a higher-dimensional context. 

For quasi-spherically symmetric states, defined by Eq. (\ref{(3+n)-D_UR_combined**}) with $\kappa_i \neq M_P/M$ for at least some $i$, the volume occupied by the particle in the $n$ extra dimensions of momentum space is
\begin{eqnarray} \label{Volume*}
V_{(n)} \sim \prod_{i=1}^{n}\Delta_{\psi}p_i \sim (cM_{P})^{n}\left(\prod_{i=1}^{n}\kappa_{i}\right)^{-1}.
\end{eqnarray}
In these states, this volume remains fixed but the total volume also depends on the 3-dimensional part $\Delta_{\psi}P$, which may take any value satisfying Eq. (\ref{(3+n)-D_UR_combined**}). The underlying physical assumption behind the mathematical requirement of fixed extra-dimensional volume is discussed further in Sec. \ref{Sec.5.2}. However, the
idea is that, since the extra-dimensional space 
can only be probed indirectly -- for example, via high-energy collisions between particles whose momenta in the compact directions cannot be directly controlled -- the net effect of any interaction is likely to leave the total extra-dimensional volume occupied by the wave packet unchanged, 
even if its 3-dimensional part can be successfully localized on scales below $R_E$. This is
the mathematical expression of the fact that we have no 
control over the extra-dimensional part of {\it any} object - including that of the apparatus used to probe a `test' higher-dimensional system. As such, complete spherical symmetry in the full higher-dimensional space cannot be guaranteed and the most natural assumption is that asymmetry persists between the 3-dimensional and extra-dimensional parts of the wave function.

Since
\begin{eqnarray}  \label{kappa_choice}
1 \leq \prod_{i=1}^{n}\kappa_{i} \leq \left(\frac{R_{E}}{R_{P}}\right)^{n},
\end{eqnarray}
we 
have
\begin{eqnarray}  \label{(3+n)-D_UR_combined***} 
R_{P}\left(\frac{cM_{P}}{\Delta_{\psi}P}\right)^{1/(1+n)} \lesssim (\Delta_{\psi}R)_{\rm min} \lesssim R_{*}\left(\frac{cM_{P}}{\Delta_{\psi}P}\right)^{1/(1+n)},
\end{eqnarray}
where $R_{*}$ is defined by Eq. (\ref{R*}). Restricting ourselves to the higher-dimensional region of the $(M,R)$ diagram,
\begin{eqnarray} \label{C}
cM_{P} \geq \Delta_{\psi}P \geq cM_E,
\end{eqnarray}
we see that, for $\Delta_{\psi}P = cM_{P}$, this gives
\begin{eqnarray} \label{B}
R_{P} \leq (\Delta_{\psi}R)_{\rm min} \leq R_*
\end{eqnarray}
for \emph{any} choice of the constants $\kappa_i$, with the extreme limits $(\Delta_{\psi}R)_{\rm min} = R_{P}$ and $(\Delta_{\psi}R)_{\rm min} = R_*$ corresponding to $\kappa_i \rightarrow 1$ and $\kappa_i \rightarrow R_E/R_{P}$, respectively. 

The first of these corresponds to the scenario $R_E \rightarrow R_* \rightarrow R_{P}$, which recovers the standard Planck length bound on the minimum radius of a Planck mass particle. In other words, if {\it all} the extra dimensions are compactified on the Planck scale, both the standard $3$-dimensional Compton and Schwarzschild formulae hold all the way down to $R_{P}$, giving the familiar intersect. However, if $R_E > R_{P}$, then $(\Delta_{\psi}R)_{\rm min}$ for a Planck mass particle is larger than the Planck length and may be as large as the critical value $R_*$. This occurs when the higher-dimensional part of the wave packet completely `fills' the extra dimensions. 

In general, the upper bound on the minimum value of $\Delta_{\psi}R$ occurs when the wave packet is space-filing in the compact directions and each of these is compacified on some scale $R_E > R_{P}$. At $\Delta_{\psi}P = cM_{P}$, this gives $(\Delta_{\psi}R)_{\rm min} = R_*$, which lies in the range $R_{P} < R_* < R_E$. For $\Delta_{\psi}P \leq cM_E$, the same scenario gives $(\Delta_{\psi}R)_{\rm min} \geq R_E$, which corresponds to the effective 3-dimensional region of the $(M,R)$ plot. In this region, the assumption of quasi-sphericity breaks down and the 3-dimensional and higher-dimensional parts of the wave packet effectively decouple with respect to measurements which are unable to probe the length/mass scales associated with the extra dimensions. We may therefore set
\begin{eqnarray}  \label{(3+n)-D_UR_combined***A} 
\Delta_{\psi}R \gtrsim R_{*}\left(\frac{cM_{P}}{\Delta_{\psi}P}\right)^{1/(1+n)}
\end{eqnarray}
as the strongest \emph{lower} bound on $\Delta_{\psi}R$, since this is the \emph{upper} bound on the value of $(\Delta_{\psi}R)_{\rm min}$. To reiterate, this comes from combining two assumptions: (a) the wave function of the particle is quasi-spherically symmetric -- in the sense of  Eq.~(\ref{Volume}) -- with respect to the \emph{entire} higher-dimensional space on scales $\Delta_{\psi}R \lesssim R_E$; and (b) wave packet is space-filling in the $n$ additional dimensions of position space.

Note that the unique 3-dimensional uncertainty relation and each of the $n$ independent uncertainty relations for the compact directions still hold individually. However, the higher-dimensional uncertainty relations are satisfied for any choice of the constants $\kappa_i$ in the range specified by Eq.~(\ref{X}) and the remaining 3-dimensional relation $\Delta_{\psi}\tilde{R} \gtrsim \hbar/\Delta_{\psi}P$ is satisfied automatically for any $\Delta_{\psi}P$ satisfying Eq.~(\ref{(3+n)-D_UR_combined**}). In the limit $\kappa_i \rightarrow R_E/R_{P}$ for all $i$, in which the wave packet completely fills the compact space, it is straightforward to verify that $(\Delta_{\psi}\tilde{R})_{\rm min} = (\Delta_{\psi}R)_{\rm min} = R_E$ when $\Delta_{\psi}P = cM_E$, so that the 3-dimensional and $(3+n)$-dimensional formulae match seamlessly. Thus, for $\Delta_{\psi}P \sim cM_{P}$, we have $(\Delta_{\psi}R)_{\rm min} \sim R_*$ but the genuine 3-dimensional radius of the wave packet is of order $(\Delta_{\psi}\tilde{R})_{\rm min} \sim R_{P}$ and, for $\Delta_{\psi}P \sim cM_E$, we have $(\Delta_{\psi}\tilde{R})_{\rm min} \sim (\Delta_{\psi}R)_{\rm min} \sim R_E$.

Under these conditions, the 3-dimensional uncertainty relation is superseded, as a lower bound on the value of $\Delta_{\psi}R$, by the limit obtained by combining \emph{all} the bounds arising from each of the separate uncertainty relations, so that 
\begin{eqnarray} \label{InequalitySeries}
\Delta_{\psi}R \geq  R_{*}\left(\frac{cM_{P}}{\Delta_{\psi}P}\right)^{1/(1+n)} \geq \frac{R_{P}M_{P}c}{\Delta_{\psi}P}.
\end{eqnarray} 
It straightforward to demonstrate that all the inequalities in this subsection hold in the range $cM_{P} \geq \Delta_{\psi}P \geq cM_E$, which corresponds to $R_* \leq (\Delta_{\psi}R)_{\rm min} \leq R_E$ in Eq. (\ref{InequalitySeries}).

\subsection{Modified Compton wavelength in higher dimensions} \label{Sec.5.2}
We now make the identifications (\ref{P_ident}), so that 
\begin{eqnarray}  \label{HD_Compton+} 
R_{C} \sim R_{E}\left(\frac{M_{P}}{M}\right)^{1/(1+n)} \sim R_{*}\left(\frac{M_{P}}{M}\right)^{1/(1+n)}.
\end{eqnarray}
This implies that, when extrapolating the usual arguments for the Compton wavelength in non-relativistic quantum theory to the case of compact extra dimensions, we should identify the \emph{geometric average} of the spread of the wave packet in $1+n$ spatial dimensions with the particle radius, but its spread in the large dimensions of momentum space with the rest mass. However, there is clearly a problem with identifying the standard deviation of the \emph{total} higher-dimensional momentum, $\mathcal{P}^2 = P^2 + p_ip^i$, with the rest mass of the particle. Since the standard deviations of the individual extra-dimensional momenta are bound from below by $\Delta_{\psi}p_i \geq cM_E$, we have  $\Delta_{\psi}\mathcal{P} \geq c M_E$. Also, pair-production is avoided for $\Delta_{\psi}\mathcal{P} \leq cM$, so $M \geq M_E$ is required for consistency. Since $R_E$ must be very small to have avoided direct detection, $M_E$ must be large and the above requirement contradicts known physics as it requires \emph{all} known particles to have masses $M > M_E$. Likewise, were we to associate $\Delta_{\psi}\mathcal{P}$ with a single length scale via the usual UP, the minimum value of the standard deviation of the total higher-dimensional position vector $(\Delta_{\psi}\mathcal{R})_{\rm min} = (\Delta_{\psi}|\vec{\mathcal{R}}|)_{\rm min}$ would be bounded from above by $(\Delta_{\psi}\mathcal{R})_{\rm min} \leq R_E$, which automatically rules out the existence of particles larger than the extra dimensions.

The manifest asymmetry of the wave packet in position space on scales less than $R_E$ (and on scales greater than $cM_E$ in momentum space) therefore \emph{requires} identifications of the form Eq. (\ref{P_ident}) in order for the standard Compton formula to hold for $R \geq R_E$ in a higher-dimensional setting. What happens to the standard formula below this scale is unclear. If the wave packet is able to adopt a genuinely spherically symmetric configuration in the full higher-dimensional space (including momentum space), then the above arguments suggest the identifications $(\Delta_{\psi}\mathcal{R})_{\rm min} \sim R_C$ and $(\Delta_{\psi}\mathcal{P})_{\rm max} \sim Mc$ for $M_E \leq M \leq M_{P}$, so that the usual Compton formula holds all the way down to $M \sim M_{P}$. The possible short-comings of this approach are that it would be valid only for spherically symmetric states and that it requires a change in the identification of the rest mass and particle radius at $M = M_E$, i.e. $R_C \sim (\Delta_{\psi}\mathcal{R})_{\rm min} \rightarrow R_C \sim (\Delta_{\psi}\tilde{R})_{\rm min}$ and $Mc \sim (\Delta_{\psi}\mathcal{P})_{\rm max} \rightarrow Mc \sim (\Delta_{\psi}P)_{\rm max}$, where $\Delta_{\psi}\tilde{R}$ and $\Delta_{\psi}P \equiv \Delta_{\psi}\tilde{P}$ denote 3-dimensional quantities, as before.

As wave packets will generally be asymmetric on scales $R \geq R_E$, it is reasonable to assume that the asymmetry will persist, even when we are able to (indirectly) probe length/energy scales associated with the extra dimensions. For example, suppose that we try to localize a particle in 3-dimensional space by constructing a spherically symmetric potential barrier. We then gradually increase the steepness of the potential well, increasing the energy and localizing the particle on ever smaller length scales. In principle, we may even shrink the 3-dimensional radius below the scale of the internal space. But what about the width of the wave packet in the compact directions? Since we did not design our initial potential to be spherically symmetric in $3+n$ spatial dimensions -- having no direct manipulative control over its form in the extra dimensions -- it is difficult to imagine that we would suddenly obtain a fully spherically symmetric potential in higher-dimensional space, simply by increasing the energy at which our `measuring device' operates. 

Similarly, we may imagine confining a particle within a spherically symmetric region of 3-dimensional space by bombarding it with photons from multiple angles. Increasing the energy of the photons then reduces the radius of the sphere. But how can we control the trajectories of the probe photons in the internal space? Since, again, in the compact space, we do not have direct manipulative control over the apparatus that creates the photons, it is impossible to ensure anything other than a random influx of photons (with random extra-dimensional momenta) in the $n$ compact directions. In this case, we would expect to be able to measure the average photon energy and to relate this to a single \emph{average} length scale, but we cannot ensure exact spherically symmetry with respect to all $3+n$ dimensions, or measure the spread of the wave packet in each individual extra dimension. 

Na{\" i}vely, we may expect the single (measured) length scale to be given by the geometric average over all $3+n$ dimensions, $[(\Delta_{\psi}\tilde{R})^3\Pi_{i=1}^{n} \Delta_{\psi}R_i]^{1/(3+n)}$, for wave packets that are spherically symmetric in the large directions but irregular in the compact space. However, the arguments proposed above suggest that the key length scale is $[\Delta_{\psi}\tilde{R} \ \Pi_{i=1}^{n} \Delta_{\psi}R_i]^{1/(1+n)}$, with only independent uncertainty relations contributing to the composite measurement. This makes sense, since it is obvious that, were we able to isolate our measurements of the 3-dimensional part of the wave packet, this would yield only a single length scale $\Delta_{\psi}\tilde{R}$; any `smearing' of this measurement due to the spread of the wave packet in the extra dimensions must be due to the $n$ additional {\it independent} widths, $\Delta_{\psi}R_i \neq \Delta_{\psi}\tilde{R}$. 

In the most extreme case, we may expect the combined effects of our experimental probing of the extra dimensions to cancel each other out, leaving the total volume of the wave packet in the compact space unchanged. This justifies Eq. (\ref{Volume*}) but does not alter the reduction of the 3-dimensional volume and, hence, of the {\it overall} volume of the wave packet, when the energy of the probe particles/potential barrier is increased. Together, these considerations lead to the scaling predicted by Eq. (\ref{HD_Compton+}).

As Eq. (\ref{HD_Compton+}) corresponds to the maximum possible asymmetry for which a single length scale can be associated with $\psi$, this should give the highest possible lower bound on the size of a quantum mechanical particle in a spacetime with $n$ compact extra dimensions. Thus, the problem of defining a single quantum radius for a particle in a semi-compactified higher-dimensional space is analogous to the problem of defining a single radius for asymmetric wave packets in the 3-dimensional case. Quantum states expected to give rise to highly asymmetric wave packets include those corresponding to spinning particles, which should form a pancake in both position and momentum space. Hence, it may be necessary, even in a 3-dimensional context, to modify the standard expression for the Compton wavelength for such states to give more than one characteristic length scale (with differing rest mass dependences), corresponding to the widths of the particle in, for example, the directions parallel and perpendicular to the spin axes. 

A related question is, how can we ascribe a wave function $\psi$ to a black hole with a Schwarzschild radius $R_S<R_E$ (cf. Casadio \cite{casadio})? In the classical theory, with only infinite dimensions, a Schwarzschild black hole is the unique spherically symmetric vacuum solution \cite{Bir23}. However, in the quantum mechanical case, our previous analysis suggests that it may be possible to associate multiple quasi-spherically symmetric wave packets with the unique classical solution, just as we can for classical (spherically symmetric) point particles. The investigation of both these points lies beyond the scope of this paper and is left for future work. 

Finally, if we interpret the Compton wavelength as marking the boundary on the $(M,R$) diagram below which pair-production rates becomes significant, we expect the presence of compact extra dimensions to affect pair-production rates at high energies. Specifically, we expect pair-production rates at energies above the (lower) mass scale associated with the compact space, $M_E \sim \hbar/(cR_E)$, to be enhanced relative to the 3-dimensional case. This is equivalent to `raising' the Compton line, i.e. decreasing its (negative) gradient on the $(M,R)$ diagram. A more detailed, fully relativistic, analysis would be needed to confirm whether this is a generic result for massive scalar fields (corresponding to uncharged matter). However, there is tentative theoretical evidence that enhanced pair-production may be a generic feature of higher-dimensional theories in which some directions are compactified (see, for example, \cite{He99,Ebetal00}) but the available literature on this subject is sparse.

To summarize our results for  higher-dimensional black holes and fundamental particles, we have  
\begin{equation}  \label{HD_Compton*} 
R_{C} \sim \left \lbrace
\begin{array}{rl}
&R_{P}\frac{M_{P}}{M} \ \ \ \ \ \ \ \ \ \ \ \ \ (R_{C} \gtrsim R_E) \\
&R_{*}\left(\frac{M_{P}}{M}\right)^{1/(1+n)} \ (R_{C} \lesssim R_E)
\end{array}\right.
\end{equation}
\begin{equation}  \label{HD_Schwarz} 
R_{S} \sim \left \lbrace
\begin{array}{rl}
&R_{P}\frac{M}{M_{P}} \ \ \ \ \ \ \ \ \ \ \ \ \ (R_{S} \gtrsim R_E) \\
&R_{*}\left(\frac{M}{M_{P}}\right)^{1/(1+n)} \ (R_{S} \lesssim R_E)
\end{array}\right.
\end{equation}
for $n$ extra dimensions compactified on a single length scale $R_E$, and these lines intersect at $(R_{*},M_{P})$, where $R_* > R_{P}$ when $R_E > R_{P}$. The crucial point is that there is no TeV quantum gravity in this scenario since the intersect of the Compton and Schwarzschild lines still occurs at $M \sim M_{P}$. The effective Planck length is reduced to $R_*$ but this does not allow the production of higher-dimensional black holes at accelerators. Thus, the constraint (\ref{nconstraint}) on the scale $R_E$ in the conventional picture no longer applies. Also, whereas the LHC probes the full higher-dimensional space in the standard case, it only probes the fourth spatial dimension (or, at most, a subset $k \leq n$ of the available extra dimensions) if the BHUP correspondence remains valid in higher dimensions.
\begin{figure}  \label{MR3}
\begin{center}
   \psfig{file=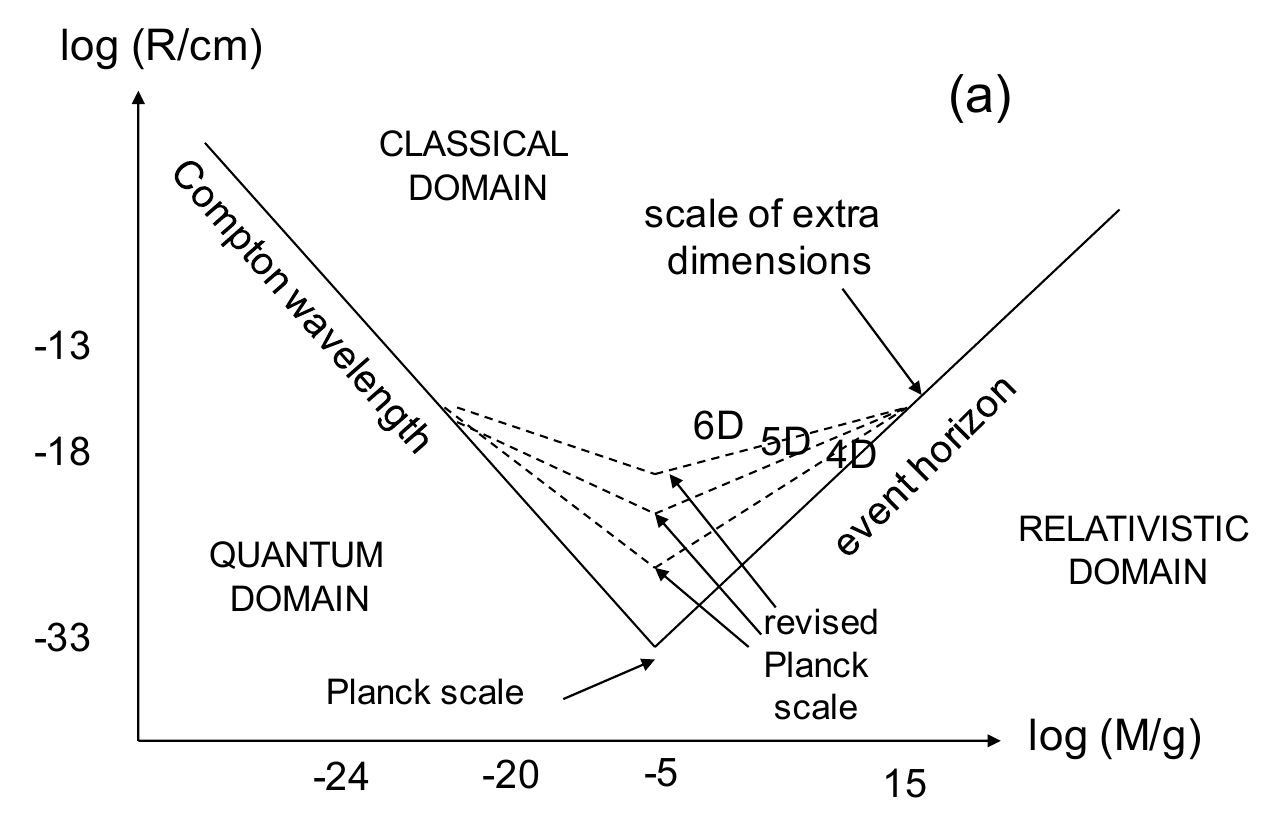,width=3.5in}
   \psfig{file=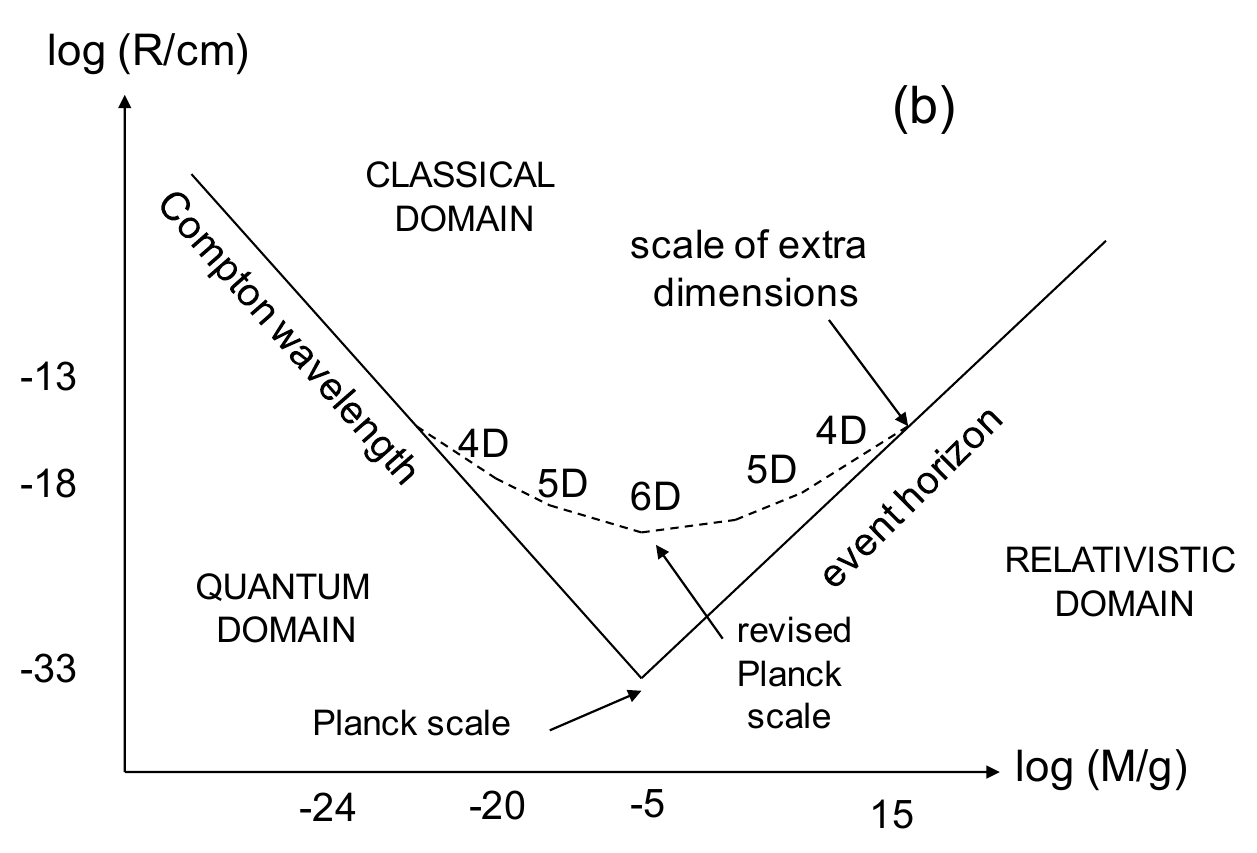,width=3.5in}
   \caption {Modifications of Fig.~2 for extra dimensions compactified on a single length scale (a) or hierarchy of length scales (b) if one imposes quasi-spherical symmetry on the higher-dimensional wave packet, preserving the duality between the Compton and Schwarzschild expressions.}
\end{center}
\end{figure}

This scenario is illustrated in Fig.~3(a) for extra dimensions compactified on a single length scale $R_E$ and in Fig.~3(b) for a hierarchy of length scales. In the latter case, the expressions (\ref{HD_Compton*})-(\ref{HD_Schwarz}) must be modified to 
\begin{equation} 
R_{C} \sim \left \lbrace
\begin{array}{rl}
&R_{P}\frac{M_{P}}{M} \ \ \ \ \ \ \ \ \ \ \ \ \ \ \ \ (R_{C} \gtrsim R_1) \\
&R_{*(k)}\left(\frac{M_{P}}{M}\right)^{1/(1+k)} \ (R_{k+1} \lesssim R_{C} \lesssim R_{k})
\end{array}\right.
\end{equation}
\begin{equation}  
R_{S} \sim \left \lbrace
\begin{array}{rl}
&R_{P}\frac{M}{M_{P}} \ \ \ \ \ \ \ \ \ \ \ \ \ \ \ \ (R_{C} \gtrsim R_1) \\
&R_{*(k)}\left(\frac{M}{M_{P}}\right)^{1/(1+k)} \ (R_{k+1} \lesssim R_{C} \lesssim R_{k})
\end{array}\right.
\end{equation}
where $R_1$ is the largest compact dimension and $R_{*(k)}$ is defined in Eq. (\ref{effectiveschwarz}). 

\section{Hawking and Compton temperatures} \label{Sec.6}

In this section, we define an `intrinsic' temperature for a fundamental particle whose mass is localized within a {\it minimum} radius $R_C$ and compare this to the Hawking temperature of a black hole, whose mass is localized within a {\it maximum} radius $R_S$. We then consider heuristic derivations of both temperatures in the 3-dimensional case using the HUP. Finally, these arguments are extended to the higher-dimensional case, using the uncertainty relations for quasi-spherically symmetric wave packets derived in Sec.~\ref{Sec.4}. 

\subsection{Hawking and Compton temperatures in three dimensions} \label{Sec.6.1}

In 3-dimensional space, we may use the usual mass-temperature relation $T \sim Mc^2/k_B$ to define the `Compton temperature' of a fundamental particle,  
\begin{eqnarray}  \label{T_C} 
T_C  \sim T_{P}\frac{M}{M_{P}} \sim T_P\frac{R_{P}}{R_C} \, ,
\end{eqnarray}
where $T_{P} \sim M_{P}c^2/k_B \sim 10^{32}$K is the Planck temperature. (In this section all equations are order of magnitude relations.) This may be interpreted as the `rest mass temperature' of a particle localized within a radius $R_C \sim R_{P} M_{P}/M$. Equation~(\ref{T_C}) is consistent with the uncertainty principle (\ref{3D_UR_SpherSymm}) together with the identifications
\begin{eqnarray}  \label{T_C_ident} 
(\Delta_{\psi}R)_{\rm min} \sim R_C \sim R_{P}\frac{T_{P}}{T_C} \, , \ \ \ (\Delta_{\psi}P)_{\rm max} \sim cM \sim cM_{P}\frac{T_{C}}{T_P} \, .
\end{eqnarray}
Hence, the Compton temperature is the \emph{maximum} temperature for a particle at rest, in the sense that the associated temperature would be lower than $T_C$ if its wave packet were localized on some scale $R > R_C$. 

This may be compared with the Hawking temperature of a black hole of mass $M$ and radius $R_S$ in three dimensions, 
\begin{eqnarray}  \label{T_H} 
T_H \sim T_{P}\frac{M_{P}}{M} \sim T_{P} \frac{R_{P}}{R_S} \, .
\end{eqnarray}
This expression is consistent with the uncertainty principle, using the relations
\begin{eqnarray}  \label{T_H_ident} 
(\Delta_{\psi}R)_{\rm max} \sim R_S \sim R_{P}\frac{T_{P}}{T_H} \, , \ \ \ (\Delta_{\psi}P)_{\rm min} \sim cM_{P}\frac{R_{P}}{R_S} \sim cM_P\frac{T_{H}}{T_P} \, .
\end{eqnarray} 
Note that here we identify the {\it maximum} possible spatial extent of the wavefunction with the Schwarzschild radius. This is based on the assumption that it is associated with a (classical) point-like mass at the central singularity which is causally disconnected from the region beyond $R_S$. It is therefore impossible to localize the mass of the black hole on scales $R > R_S$, since the horizon is a global feature of spacetime. This may be contrasted with the fundamental particle case, in which the mass cannot be localized on scales $R < R_C$. Hence $T_H$ may be interpreted as the \emph{minimum} temperature for a black hole from the perspective of an external observer, in the sense that, were the mass of the black hole to be localized on a smaller scale (e.g. due to a fluctuation in the horizon size), the temperature of the horizon would increase. 

Clearly, the Hawking and Compton temperatures coincide at $M \sim M_{P}$, giving $T_C \sim T_H \sim T_{P}$. In general, they are dual under the transformation
\begin{eqnarray}  \label{TempDual} 
T_H \sim \frac{T_{P}^2}{T_C}  \, .
\end{eqnarray}
In this analysis, we assume that the mass scale $M_P$ marks a division between elementary particles and a black holes, so that the expressions for the Compton and Hawking temperature only apply for $M < M_P$ and $M > M_P$, respectively. In a scenario in which there is no such division, so that elementary particles can be interpreted as sub-Planckian black holes, $T_C$ would itself be interpreted as a Hawking temperature \cite{CaMoPr:2011}. 

\subsection{Hawking and Compton temperatures in higher dimensions} \label{Sec.6.2}

The situation changes radically in the higher-dimensional case. If all the extra dimensions have the same compactification scale $R_E$, then the Hawking temperature is modified to \cite{CaDaMa03,CaDa04} 
\begin{eqnarray}  \label{T_H_HD} 
T_H \sim \frac{M_{P}'c^2}{k_B}\left(\frac{M_{P}'}{M}\right)^{1/(1+n)} \sim T_*\left(\frac{M_{P}}{M}\right)^{1/(1+n)} .
\end{eqnarray}
Here $M_P'$ is the $(3+n)$-dimensional Planck mass, 
given by Eq. (\ref{revisedplanck}), and the second relationship follows from the definitions
\begin{eqnarray}  \label{T*} 
T_*  \equiv (T_PT_E^n)^{1/(1+n)}, \quad T_E \equiv \frac{M_Ec^2}{k_B}  \sim T_{P}\frac{R_{P}}{R_E} \, .
\end{eqnarray}
The higher-dimensional Hawking temperature line now intersects the 3-dimensional Compton temperature Eq.~(\ref{T_C}) line at $M \sim M_{P}'$, giving 
\begin{eqnarray}  \label{T_P'} 
T_{P}' \sim (T_{P}^2T_E^n)^{1/(2+n)} \sim c^2 M_{P}'/k_B
\end{eqnarray}
as the revised Planck temperature. In Refs.~\cite{CaDaMa03,CaDa04}, Eq.~(\ref{T_H_HD}) was derived from the relations
\begin{eqnarray}  \label{T_H_ident+} 
(\Delta_{\psi}R)_{\rm max} \sim R_S 
\sim R_{P}\frac{T_{P}}{T_H}, \ \ \ (\Delta_{\psi}P)_{\rm min} \sim cM_{P}\frac{R_{P}}{R_S} \sim cM_*\frac{T_H}{T_*} \, ,
\end{eqnarray}
where $M_* \sim \hbar/(cR_*)$, which reduce to Eq.~(\ref{T_H_ident}) for $n=0$. The $M$ dependence could also be obtained from the surface gravity, since this scales as  $M/R_S^{2+n} \propto M^{1/(1+n)}$.

As discussed in Sec. \ref{Sec.5}, we expect the usual 3-dimensional uncertainty principle to hold all the way to the Planck point $(M_{P},R_{P})$ for particle wave packets that are spherically symmetric with respect to all $3+n$ spatial dimensions. For black hole wave functions that are totally spherically symmetric in the $(3+n)$-dimensional space on scales $R_S \lesssim R_E$, one might therefore expect standard identifications 
like (\ref{T_H_ident+}) 
to be valid if one makes 
the substitutions $(\Delta_{\psi}R)_{\rm min} \rightarrow (\Delta_{\psi}\mathcal{R})_{\rm min}$, $(\Delta_{\psi}P)_{\rm min} \rightarrow (\Delta_{\psi}\mathcal{P})_{\rm min}$, where $\Delta_{\psi}\mathcal{R}$ and $\Delta_{\psi}\mathcal{P}$ denote the uncertainties in the \emph{total} $(3+n)$-dimensional position and momentum, respectively.
However, in Sec. \ref{Sec.5.2}, the total spherical symmetry of the wave packet in $3+n$ dimensions led us to identify the particle rest mass with the maximum value of the uncertainty for the total higher-dimensional momentum, $(\Delta_{\psi}\mathcal{P})_{\rm max}$.
This resulted in inconsistencies (i.e. to  all known particles having masses $M \geq M_E$ and radii $R_C \leq R_E$). Likewise, the total spherical symmetry of the black hole wave packet in $3+n$ dimensions suggests the identification $(\Delta_{\psi}\mathcal{P})_{\rm min} \sim cM_{P}^2/M$, which implies that all black holes should have $M \leq M_E'$ and $R_S \leq R_E$, thus rendering the $(3+1)$-dimensional part of the $(M,R)$ diagram inaccessible. 

The solution (as before) is to consider seriously the manifest asymmetry between the 3-dimensional and extra-dimensional parts of the wave packet and to combine the individual uncertainty relations for each dimension to obtain Eq.~(\ref{(3+n)-D_UR_combined***A}) for quasi-spherically symmetric systems. Applying this to `particles' with masses $M > M_{P}$ (i.e. black holes) leads us to identify $(\Delta_{\psi}P)_{\rm min} \sim cM_{P}^2/M$ with the \emph{3-dimensional} part of the wave function. Clearly, this gives the standard expression (\ref{higherBH}) for $R_S$ in $(3+n+1)$-dimensional spacetime. Then the identification $R_S \sim R_{P}T_{P}/T_H$ recovers the expression for the higher-dimensional Hawking temperature derived in Refs.~\cite{CaDaMa03,CaDa04}.

For fundamental particles in higher-dimensional space, this suggests the identifications 
\begin{eqnarray}  \label{T_C_ident} 
(\Delta_{\psi}R)_{\rm min} \sim R_C \sim R_*\left(\frac{M_{P}}{M}\right)^{1/(1+n)} , \ \ \ (\Delta_{\psi}P)_{\rm max} \sim cM_{P}\left(\frac{R_*}{R_C}\right)^{1+n} \sim cM \, ,
\end{eqnarray}
which reduces to Eq.~(\ref{T_C_ident}) for $n=0$. Together with $R_C \sim R_{P}T_{P}/T_C$, this leads to
\begin{eqnarray}  \label{x} 
T_C \sim T_*\left(\frac{M}{M_{P}}\right)^{1/(1+n)}.
\end{eqnarray}
This is the only definition which is consistent with our previous argument that the particle rest mass should be identified with the 3-dimensional part of the momentum spread for a quasi-spherical symmetric wave packet, so that Eqs. (\ref{T_C_ident}), (\ref{x}) and (\ref{(3+n)-D_UR_combined***A}) are self-consistent.

In this scenario, the lines corresponding to the higher-dimensional Compton and Hawking temperatures in the $(M,R)$ diagram now intersect at $M \sim M_{P}$, giving $T_C \sim T_H \sim T_*$. Physically, we may interpret this as implying that, even with compact extra dimensions, evaporating black holes form Planck mass relics with radii of order $R_C \sim R_S \sim R_* > R_{P}'$ and temperatures $T_* < T_{P}'$. This suggests that, in the quantum mechanical description, attempts to either increase or decrease the energy of the relics, always increase the radius of the corresponding wave packet above $R_*$, while the temperature always increases. The crucial difference is that particle/anti-particle pairs are absorbed in the former case, creating a Schwarzschild black hole, whereas, in the latter, mass-energy is carried away by the pair-production of particles and the relic itself becomes a particle, with no event horizon. Furthermore, these actions are symmetric, in that a mass shift of fixed magnitude results in corresponding (fixed) changes in radius and temperature. 
Alternatively, our picture suggests that attempts to localize the relic wave packet on scales $\Delta_{\psi}R < R_*$ will result in pair-production of fundamental particles, whereas attempts to localize it on scales $\Delta_{\psi}R > R_*$ lead to the production of microscopic black holes. In other words, we can multiply particles by squashing them and black holes by pulling them apart.

Although the results obtained above assume that
the extra dimensions are compactified on a single cale $R_E$, we note that similar arguments apply when there is a hierarchy of compactification scales. In this case $T_*$, defined in Eq. (\ref{T*}), is replaced by $T_{*(k)} \sim \left(T_P\prod_{i=1}^{n}T_i\right)^{1/(1+k)}$, where $T_i \sim M_i c^2/k_B$, in all relevant formulae, which are then valid for $R_{k+1} \lesssim R \lesssim R_{k}$, or the equivalent mass range, as before. 

\section{Observational consequences} \label{Sec.7}

In this section we consider two possible observational consequences of retaining the duality between the Compton and Schwarzschild expressions. The first relates to the detectability of exploding primordial black holes (PBHs). There is still no unambiguous detection of such explosions but it has been claimed that some short-period gamma-ray bursts  could be attributed to PBHs \cite{cline}.  
The second relates to high-energy scattering experments and the enhancement of pair-production at at accelerators on scales below $R_E$. 

\subsection{Black hole explosions}  \label{Sec.7.1}

In the standard ($n=0$) model, PBHs complete their evaporation at the present epoch if they have an initial $M_0 \approx 5 \times 10^{14}$g and an initial radius $R_0 \approx 10^{-13}$cm, comparable to the size of a proton \cite{cksy}. For most of their lifetime these PBHs  are producing photons with energy $E_0 \approx 100$~MeV, so the extragalactic $\gamma$-ray background at this energy places strong constraints on the number of such PBHs and thereby their current explosion rate. In principle, these PBHs could also contribute to cosmic-ray positrons and antiprotons, although there are other possible sources of these particles \cite{cksy}.

If there are $n$ extra dimensions, each with compactification scale $R_E$, then the mass and temperature of a PBH terminating its evaporation today will change if the critical radius $R_0$ is less than $R_E$. 
From Eq.~(\ref{T_H_HD}) and the higher-dimensional black-body formula, the mass loss rate should then be
\begin{eqnarray} 
dM/dt \propto R^2 T^{4+n} \propto M^{-(2+n)/(1+n)} \, ,
\end{eqnarray} 
leading to a black hole lifetime
\begin{eqnarray} 
\tau \sim  M/(dM/dt) \propto M^{(3+2n)/(1+n)} \, .
\end{eqnarray} 
Thus the critical mass of the PBHs evaporating at the present epoch and the associated temperaure become
\begin{eqnarray} 
M_0 \propto t_0^{(1+n)/(3+2n)} \, , \quad T_0 \propto t_0^{-1/(3+2n)} \, ,
\end{eqnarray} 
so both the critical mass and the associated temperature are increased compared to the $3$-dimensional case ($n=0$). If there is a hierarchy of extra dimensions, the value of $n$ in the above equations must be replaced by $n_k$, which is the  dimensionality for which $R_{k+1} < R_0 < R_k$ ($k \leq n$). This means that the standard limits on their number density must also be changed, though we do not discuss this further here.

An important point from an observational perspective is that the black holes evaporating at the present epoch are necessarily higher dimensional if $R_E > 10^{-13}$cm. In the TeV quantum gravity scenario, for example, Eq.~(\ref{nconstraint}) implies $R_E > 10^{-13}$cm
providing $n< 7$. This condition is necessarily satisfied in M-theory because the maximum number of compactified dimensions is $7$. Figure~4 shows the $(M,R)$ diagram for a scenario in which there are three compactified dimensions. 
In the case of compactified extra dimensions, another uncertainty arises due to the fact that quanta emitted in the compact directions may simply be reabsorbed due to the periodic boundary conditions. 

\begin{figure}  \label{MR3}
\begin{center}
     \psfig{file=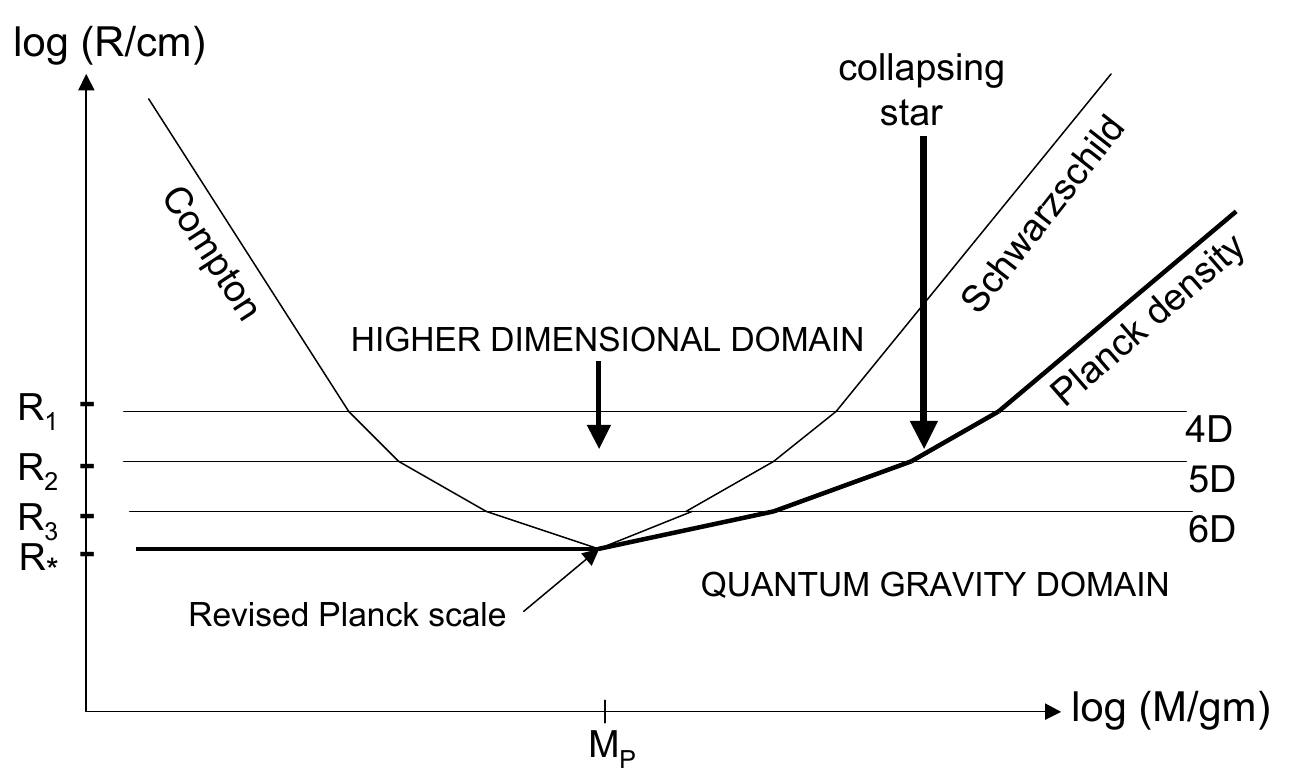,width=3.5in}
     \caption {Showing classical, quantum, relativistic and higher-dimensional ($M,R$) domains for a model with three compactified dimensions in which the Compton-Schwarzschild duality is preserved.}
\end{center}
\end{figure}


\subsection{Consistency with $D-$particle scattering results}  \label{Sec.7.2}

We now consider the consistency of our phenomenologically general results with respect to the leading higher-dimensional theory of fundamental physics: string theory. In particular, we focus on their consistency with minimum-radius results for higher-dimensional, non-relativistic and quantum mechanical particle-like objects, known as $D$-particles. 
The end points of open strings obey Neumann or Dirichlet boundary conditions (or a combination of both) and are restricted to $(p+1)$-dimensional submanifolds, where $p \leq 3+n$, called $D_p$-branes. Although these are composite rather than  fundamental objects, they have dynamics in their own right and an intrinsic tension $\mathcal{T}_p = (g_sl_s^{p+1})^{-1}$, where $g_s$ denotes the string coupling and $l_s$ is the fundamental string length scale \cite{Hossenfelder:2012jw}. Thus, $D_0$-branes, also referred to as $D$-particles, are point-like, and possess internal structure only on scales $\lesssim g_sl_s$. This may be seen as the analogue of the Compton wavelength in $D_0$-brane models of fundamental particles.

At high energies, strings can convert kinetic into potential energy, thereby increasing their extension and counteracting attempts to probe smaller distances. Therefore, the best way to probe $D_p$-branes is by scattering them off each other, instead of using fundamental strings as probes \cite{Bachas:1998rg}. $D$-particle scattering has been studied in detail by Douglas {\it et al} \cite{Douglas:1996yp}, who showed that slow moving $D$-particles can be used to probe distances down to $g_s^{1/3}l_s$ in $D=10$ spacetime dimensions. 

This result may be obtained heuristically as follows \cite{Hossenfelder:2012jw}. Let us consider a perturbation of the metric component $g_{00} = 1+2V$ induced by the Newtonian potential $V$ of a higher-dimensional particle of mass $M$. In $D$ spacetime dimensions, this takes the form
\begin{eqnarray} \label{Dpart-1}
V \sim -  \frac{G_DM}{(\Delta x)^{D-3}} \, ,
\end{eqnarray}
where $\Delta x$ is the spatial extension of the particle and $G_D$ is the $D$-dimensional Newton's constant, so that the horizon is located at
\begin{eqnarray} \label{Dpart-2}
\Delta x \sim (G_DM)^{1/(D-3)}. 
\end{eqnarray}
(For convenience, we set $c=\hbar=1$ throughout this section.) In spacetimes with $n$ compact spatial dimensions, this is related to the $(3+1)$-dimensional Newton's constant via $G \sim G_D/R_E^n$, so that, for $D = 3+n+1$, we simply recover the formula for the higher-dimensional Schwarzschild radius (\ref{higherBH}).

However, we may also use Eq.~(\ref{Dpart-2}) to derive the minimum length obtained from $D$-particle scattering in \cite{Douglas:1996yp} by first setting $M \sim \hbar/\Delta t$, where $\Delta t$ is the time taken to test the geometry, and instead using the higher-dimensional Newton's constant derived from string theory, $G_D \sim g_s^2 l_s^{D-2}$ \cite{Zwiebach:2004tj}. This gives
\begin{eqnarray} \label{Dpart-3}
(\Delta t)(\Delta x)^{D-3} \gtrsim g_s^2l_s^{D-2} \, .
\end{eqnarray}
Combining this with the spacetime uncertainty principle, which is thought to arise as a consequence of the conformal symmetry of the $D_p$-brane world-volume \cite{Yoneya:1989ai,Yoneya:2000bt,Jevicki:1998yr},
\begin{eqnarray} \label{Dpart-4}
\Delta x \Delta t \gtrsim l_s^2,
\end{eqnarray}
we then have
\begin{eqnarray} \label{Dpart-5}
(\Delta x)_{\rm min} \sim g_s^{2/(D-4)}l_s, \ \ \ (\Delta t)_{\rm min} \sim g_s^{-2/(D-4)}l_s.
\end{eqnarray}
For $D=10$, this gives $(\Delta x)_{\rm min} \sim g_s^{1/3}l_s$, as claimed.
 
Combining results from string theory and higher-dimensional general relativity by setting $G_D \sim g_s^2l_s^{D-2} \sim R_{P}^2R_E^{D-4}$ with $D=3+n+1$, we obtain
\begin{eqnarray} \label{Dpart-6}
R_{P}' \sim (R_{P}^2R_E^n)^{1/(2+n)} \sim g_s^{2/(2+n)}l_s \, ,
\end{eqnarray}
which gives $R_{P}' \sim g_s^{1/4}l_s$ as the modified Planck length for $D = 10$. In fact, Eqs. (\ref{Dpart-5})-(\ref{Dpart-6}) suggest that the minimum positional uncertainty for $D$-particles {\it cannot} be identified with the modified Planck length obtained from the intersection of the higher-dimensional Schwarzschild line, $R_S \sim M^{1/(1+n)}$, and the standard Compton line, $R_C \sim M^{-1}$, in any number of dimensions. Hence, the standard scenario is {\it incompatible} with $D$-particle scattering results.

However, it is straightforward to verify that, if $R_{P} \sim g_s^{-2/n}l_s$, we have $R_* \sim (R_{P}R_E^n)^{1/(1+n)} \sim (\Delta x)_{\rm min} \sim g_s^{2/n}l_s$, so that the intersection of the higher-dimensional Schwarzschild and Compton lines is equal to the minimum length scale that can be probed by $D$-particles. In this scenario, $R_E \sim g_s^{2(1+n)/n^2}l_s$, and we note that $R_E \rightarrow R_* \rightarrow R_{P}' \rightarrow R_{P} \rightarrow l_s$ for $g_s \rightarrow 1$, as required for consistency, but, in general, $R_* > R_{P}'$ (or equivalently $R_E > R_{P}$), requires $g_s > 1$.

\section{Conclusions} \label{Sec.8} 

We have addressed the question of how the effective Compton wavelength of a fundamental particle -- defined as the minimum possible positional uncertainty over measurements in {\it all} dimensions -- scales with mass if there exist $n$ extra compact spatial directions. In $(3+1)$-dimensional spacetime, the Compton wavelength scales as $R_C \sim M^{-1}$, whereas the Schwarzschild radius scales as $R_S \sim M$, so the two are related via $R_S \sim R_{P}^2/R_C$. In higher-dimensional spacetimes with $n$ compact extra dimensions, $R_S \sim M^{1/(1+n)}$ on scales smaller than the compactification radius $R_E$, which breaks the symmetry between particles and black holes if the Compton scale remains unchanged. However, we have argued that the effective Compton scale, defined in terms of minimum positional uncertainty, depends on the form of the wavefunction in the higher-dimensional space. If this is spherically symmetric in the three large dimensions, but maximally asymmetric (i.e. pancaked) in the full $3+n$ spatial dimensions, then the effective radius scales as $R_C \sim M^{-1/(1+n)}$ rather than $M^{-1}$ on scales less than $R_E$ and this preserves the symmetry about the $M \sim M_{P}$ line in ($M,R$) space. 

In this scenario, the effective Planck length is reduced but the Planck mass is unchanged, so quantum gravity and microscopic black hole production are associated with the standard Planck energy, as in the 3-dimensional scenario. On the other hand, one has the interesting prediction that the Compton line -- which marks the onset of pair-production -- is `lifted', relative to the 3-dimensional case, in the range $R_{P} < R < R_E$, so that extra-dimensional effects may become visible via enhanced pair-production rates for particles with energies $E > M_Ec^2 = \hbar c/R_E$. This prediction may be consistent with minimum length uncertainty relations obtained from $D$-particle scattering amplitudes in string theory.  
Also, as indicated in Fig.~4, the existence of extra compact dimensions has crucial implications for the detectability of black holes evaporating at the present epoch. Since these have the size of a proton, they are {\it necessarily} higher-dimensional for $R_E > 10^{-13}$cm.

Our results naturally suggest a definition for the intrinsic `quantum' temperature of a fundamental particle, here referred to as the `Compton temperature', which is associated with the spatial localization of a particle's wave packet within a minimum (i.e. Compton) radius. This scales as $T_C \sim M^{1/(1+n)}$ in ($3+n+1$)-dimensional spacetime. Using the dimensional dependence of 
$R_C$ and $R_S$, we find that this is related to the Hawking temperature $T_H$ of a black hole via $T_C \sim T_{P}^2/T_H$ 
in arbitrary dimensions, so that a particle of mass $M < M_{P}$ has the same temperature as a black hole with dual mass $M' = M_{P}^2/M$.

\if
The above analysis assumes that the Compton and Schwarzschild lines retain simple power-law forms until their intersection point in Figs. 1 or 2(a)-(b). 
No allowance is made for deviations from the HUP, as postulated by the GUP, and no attempt is made to smooth the transition at the Planck point, as postulated by the BHUP correspondence. These effects would obviously entail different temperature predictions in the Planck regime even in the $3$-dimensional case. Our main intention is to examine the consequences of the above analysis for the temperature predicted by the BHUP correspondence in the higher-dimensional case. 
\fi

In this paper, we have assumed that non-relativistic quantum mechanical particles obey the standard Heisenberg Uncertainty Principle (HUP) in each spatial direction. The modified expression for the {\it effective} Compton line, which retains a simple power-law form until its intersection with the higher-dimensional Schwarzschild line in the ($M,R$) diagram, is seen to arise from the application of the HUP to maximally asymmetric wave functions. These are spherically symmetric with respect to the three large dimensions but pancaked in the compact directions. No allowance has been made for deviations from the HUP, as postulated by various forms of Generalised Uncertainty Principle (GUP), proposed in the quantum gravity literature, and no attempt has been made to smooth out the transition between particle and black hole states at the Planck point, as postulated by the Black Hole Uncertainty Principle (BHUP) correspondence \cite{Ca:2014}. As discussed in a separate paper \cite{paper3}, these effects would entail different temperature predictions in the Planck regime even in the $3$-dimensional case. Our main intention here has been to examine the consequences of the existence of extra dimensions in the `standard' (i.e. HUP-based) scenario.

\begin{center}
{\bf Acknowledgments}
\end{center}

We thank the Research Center for the Early Universe (RESCEU), University of Tokyo, for hospitlaity received during this work. 
We thank Pichet Vanichchapongjaroen, Shingo Takeuchi and Tiberiu Harko for helpful discussions during the preparation of the manuscript. 

\appendix

\section{A more general classification of physical systems } \label{AppendixA}

We have three dichotomies for the physical systems discussed in this paper: classical/quantum, non-relativistic/relativistic and weak-gravitational/strong-gravitational. This gives $8$ possible regimes, according to the values of different combinations of the characteristic mass $(M)$, length $(R)$ and time $(t)$ scales of the system considered. These are shown below, together with examples of physical formulae from the corresponding regime:
\begin{subequations} 
\begin{align}
\frac{MR^2}{t} &\gg \hbar \implies {\rm classical}  \ \ \  (\Delta_{\psi}x\Delta_{\psi}p \gg \hbar)
\label{Class} 
\\
\frac{MR^2}{t} & \sim \hbar \implies {\rm quantum}  \ \ \ (\Delta_{\psi}x\Delta_{\psi}p \sim \hbar) 
\label{Quant}
\end{align}
\end{subequations} 
\begin{subequations} 
\begin{align}
\frac{R}{t} &\ll c \implies {\rm non-relativisitic}  \ \ \ (v \ll c), \label{Non-rel} \\
\frac{R}{t} &\sim c \implies {\rm relativistic}  \ \ \ \ \ \ \ \ \ \ \ \ \ (v \sim c), \label{Rel}
\end{align}
\end{subequations}
\begin{subequations} 
\begin{align}
\frac{R^3}{t^2M} &\gg G \implies {\rm weak-gravitational}  \ \ \ \ (R \gg R_S = 2GM/c^2, \ G\mu \ll c^2), \label{Non-grav} \\
\frac{R^3}{t^2M} &\sim G \implies {\rm strong-gravitational}  \ \ \ (R \sim R_S = 2GM/c^2, \ G\mu \sim c^2). \label{Grav}
\end{align}
\end{subequations}
Systems that fall within the ranges described by Eqs.~(\ref{Class}), (\ref{Non-rel}) and (\ref{Non-grav}) can be classified as classical, non-relativistic and weakly-gravitational. These include all systems described by Newton's laws in the absence of gravity. Systems described by Newtonian gravity fall into the ranges specified by Eqs.  (\ref{Class}), (\ref{Non-rel}) and (\ref{Grav}), etc. 

Systems for which $\Delta_{\psi}x\Delta_{\psi}p \gg \hbar$ can be adequately described by classical equations of motion via the correspondence principle \cite{Rae00,VanVleck:1928zz,Got98} and we can ignore relativistic effects in systems with $v \ll c$. The formulae on the right-hand side of Eq.~(\ref{Non-grav}) refer to spherically symmetric bodies of mass $M$ and to cosmic strings of mass per unit length $\mu$. For $G\mu \ll c^2$, the deficit angle of the spacetime surrounding the string core is small \cite{Vi81} and for an observer at a distances $R \gg R_S \sim GM/c^2$ from a spherically symmetric body, its gravitational field may be ignored. Fundamental particles typically fall into this category, since our observations are limited to the region $R > R_C \gg R_S$, for $M \ll M_{P}$. Hence, these may be described by quantum mechanics if they fall into the regimes specified by Eqs. (\ref{Quant}) and (\ref{Non-rel}) or by quantum field theory if they fall into those specified  by Eqs. (\ref{Quant}) and (\ref{Rel}). Quantum gravity applies to systems lying within the ranges specified by Eqs. (\ref{Quant}), (\ref{Rel}) and (\ref{Grav}). To adequately illustrate each of the eight regimes, we would need to combine the $(M,R)$ diagrams used throughout this paper with a $t$-axis to give a 3-dimensional representation. 

In light of the discussion above, the Compton line may be seen as marking the boundary in the $(M,R)$ plane between the quantum$-$non-relativistic$-$weak-gravitational and the quantum$-$relativistic$-$weak-gravitational regimes, whereas the Schwarzschild line marks the boundary between regions corresponding to the classical$-$non-relativisitic$-$strong-gravitational and classical$-$relativistic$-$strong-gravitational regimes. The exception, for both lines, is the area around the point of intersection, close to the Planck point $(M_{P},R_{P})$, in which we expect quantum$-$relativisitic$-$strong-gravitational effects to become important. 


\end{document}